\documentclass[%
 reprint,
 amsmath,amssymb,
 aps,
]{revtex4-1}

\usepackage{graphicx}
\usepackage[space]{grffile}
\usepackage{latexsym}
\usepackage{textcomp}
\usepackage{longtable}
\usepackage{tabulary}
\usepackage{booktabs,array,multirow}
\usepackage{amsfonts,amsmath,amssymb}
\usepackage{natbib}
\usepackage{url}
\usepackage{hyperref}
\hypersetup{colorlinks=false,pdfborder={0 0 0}}
\usepackage{etoolbox}

\newif\iflatexml\latexmlfalse

\AtBeginDocument{\DeclareGraphicsExtensions{.pdf,.PDF,.eps,.EPS,.png,.PNG,.tif,.TIF,.jpg,.JPG,.jpeg,.JPEG}}

\usepackage[utf8]{inputenc}
\usepackage[ngerman,english]{babel}

\begin{document}

\title{Experimental evidence of non-classical  brain functions}

\author{Christian Kerskens}
\affiliation{Institute of Neuroscience, Trinity College, Dublin, Ireland}

\author{David López Pérez}
\affiliation{Institute of Neuroscience, Trinity College, Dublin, Ireland}

\selectlanguage{english}
\begin{abstract}
Recent proposals in quantum gravity have suggested that unknown systems can mediate entanglement between two known quantum systems, if and only if the mediator itself is non-classical. 
\\
This approach may be applicable to the  brain, where  speculations about quantum operations in consciousness and cognition have a long history. \\
Proton spins of bulk water, which most likely  interfere with any brain function, can act as the known quantum systems. If an unknown mediator exists, then NMR methods based on multiple quantum coherence (MQC) can act as entanglement witness.   
However, there are doubts that today's NMR signals can contain quantum correlations in general, and specifically in the brain environment.\\
Here, we used a witness protocol based on zero quantum coherence (ZQC) whereby we minimised the classical signals to circumvent the NMR detection limits for quantum correlation. 
\\
For short repetitive periods, we found evoked signals  in most parts of the brain, whereby the temporal appearance resembled heartbeat-evoked potentials (HEPs). We found that those signals had no correlates with any classical NMR contrast.
Similar to  HEPs, the evoked signal depended on conscious awareness. Consciousness-related or electrophysiological signals are  unknown in NMR. 
Remarkably, these signals only appeared if the local properties of the magnetisation were reduced. \\
Our findings suggest that we may have  witnessed entanglement  mediated by  consciousness-related brain functions. Those brain functions  must then operate non-classically, which would mean that consciousness is non-classical. 
\end{abstract}%

\maketitle

\section{Introduction}\label{sec1}
Quantum mechanisms are at work in sensory systems feeding the brain with information \cite{doi:10.1126/science.1925597,Keller:2004uh,doi:10.1146/annurev-biophys-032116-094545}. Foremost in  magneto-reception \cite{wiltschko1968}, there is no doubt that only quantum mechanical effects can explain its sensitivity \cite{doi:10.1146/annurev-biophys-032116-094545}. It has been suggested that entangled radical electron pairs are involved.  
\\
Beyond those sensory inputs, more complex brain functionalities depend on the presence of specific nuclear spins. For example, Lithium-6 isotopes with nuclear spin 1  increase activity of complex behaviour in contrast to  Lithium-7 isotopes with 3/2 spin where it decreases \cite{SECHZER19861258}. Similar, Xenon isotopes with 1/2 spin are effective anaesthetizers in contrast to Xenon isotopes with spin 0 which have only little effects \cite{Li_2018}. 
\\
However, nuclear spins can, like electron spins, influence chemical reactions \cite{Steiner:1989tt}, which then lead to macroscopic results as commonly observed in physiology. Whether those or other macroscopic systems in the brain can be non-classical, is still unknown. Experimental methods, which could distinguish classical from quantum correlations in the living brain, haven't yet been established.  
\\
In this respect, recent proposals in quantum gravity \cite{PhysRevLett.119.240402,PhysRevLett.119.240401} may help to overcome experimental restrictions in living systems. Those proposals use auxiliary quantum systems for which  they showed that if a system can mediate entanglement between auxiliary quantum systems then the mediator itself is non-classical. 
\\
If a cerebral mediator of this kind exist, then it is likely that the entanglement plays an important role in the brain.
Although, quantum computing can be achieved without entanglement \cite{BIHAM200415}, it is commonly believed that entanglement is essential to play out its full advantages \cite{BIHAM200415}. Therefore, it is likely that entanglement, if mediated by any brain function at all, may only occur during brain activity. 
\\
Hence, the experimental demands on an auxiliary quantum system are that they can be  measured non-invasively in the conscious-aware brain, and further that entanglement can be witnessed. \\
A non-invasive approach offers NMR. The nuclear spins are quantum systems which could, in theory, be entangled by a cerebral mediator. NMR sequences based on multiple quantum coherence (MQC) are also able to witness entanglement \cite{PhysRevLett.120.040402}.
\\
The MQC entanglement witness relies on bounds which, for applications in biology, may be based on the maximal classical signal achievable. The maximal classical MQC signal in fluids have been estimated on the basis of the intermolecular MQC (iMQC) approach \cite{225062:5067241}. The iMQC signal, despite the naming, is an entirely classical signal because it can also be the classically derived \cite{Jeener_2000} which is known as multiple spin echo (MSE)\cite{225062:5039968,Bowtell1990}. Therefore, it can be used as the classical bound. 
\\
Further, an exclusion of classicality can also be argued on the following basis. A single quantum coherence (SQC) which is weighted by susceptibility or diffusion contrast may respond similar to physiological changes as the iMQC contrast which is caused by long-range dependency or  rotational symmetry breaking \cite{225062:5067412,225062:5041128,225062:5067413,225062:5067414}. Hence, a signal change in a MQC sequence with no corresponding diffusion or T$_2^*$-weighted SQC signal  is most likely non-classically.   
With this knowledge at hand, we can now search for situations in which witnessing entanglement may be possible. As mentioned before brain activity, or more concretely brain computation, may play a crucial role in the creation of cerebral entanglement. 
Hence, we can make additional observations specific to the brain. We propose the following conditions:
\\
(1) Sufficient condition for witness -- Non-invasive direct detection of brain computation is possible using electrophysiological measurements. With today's technology, corresponding MRI signals are undetectable. Therefore we conclude that a detection of an electrophysiological event with a conventional MRI system, which classically is not possible, would be sufficient to witness entanglement. 
\\
(2) Necessary condition for witness -- The brain is able to operate without any external magnetic fields which means that, without a brain function at work, all states are initially mixed. Hence, the assumed brain function producing entanglement, must use a kind of quantum distillation process \cite{PhysRevA.80.041603} on mixed states \cite{PhysRevA.71.022316}. Therefore, we conclude that the NMR signal must initially be saturated.
\\
The following two arguments underpin the importance of saturation for the detection  further.
The unusual dispense of (pseudo-)pure states on which the MR signal is normally constituted, circumnavigates the major problem that entanglement of pure spins, which are in close proximity, is highly unlikely \cite{PhysRevLett.83.1054}. Further, the saturation of pure local states may serve the existence of non-localities because local and non-local properties can be complimentary \cite{10.1088/1367-2630/7/1/258,Fan_2019}. 
\\
Now, we are in the position to address the question whether the brain can mediate entanglement, experimentally.
Based on the above considerations, we explored if the conscious-aware brain may use entanglement during computing. As indicators of brain computation, we focussed on electrophysiological brain waves, which can be observed in the conscious-aware brain at rest. 
\\
We acquired  MRI time series which were highly saturated and which were able to detect ZQC. Based on the maximal temporal resolution of our method ($<$ 5Hz), we focussed on Heartbeat Evoked Potentials (HEPs) \cite{PARK2019502} which like other electrophysiological signals are far below the detection threshold of conventional MRI sequences. 
\begin{figure}[h]
\begin{center}
\includegraphics[width=1.00\columnwidth]{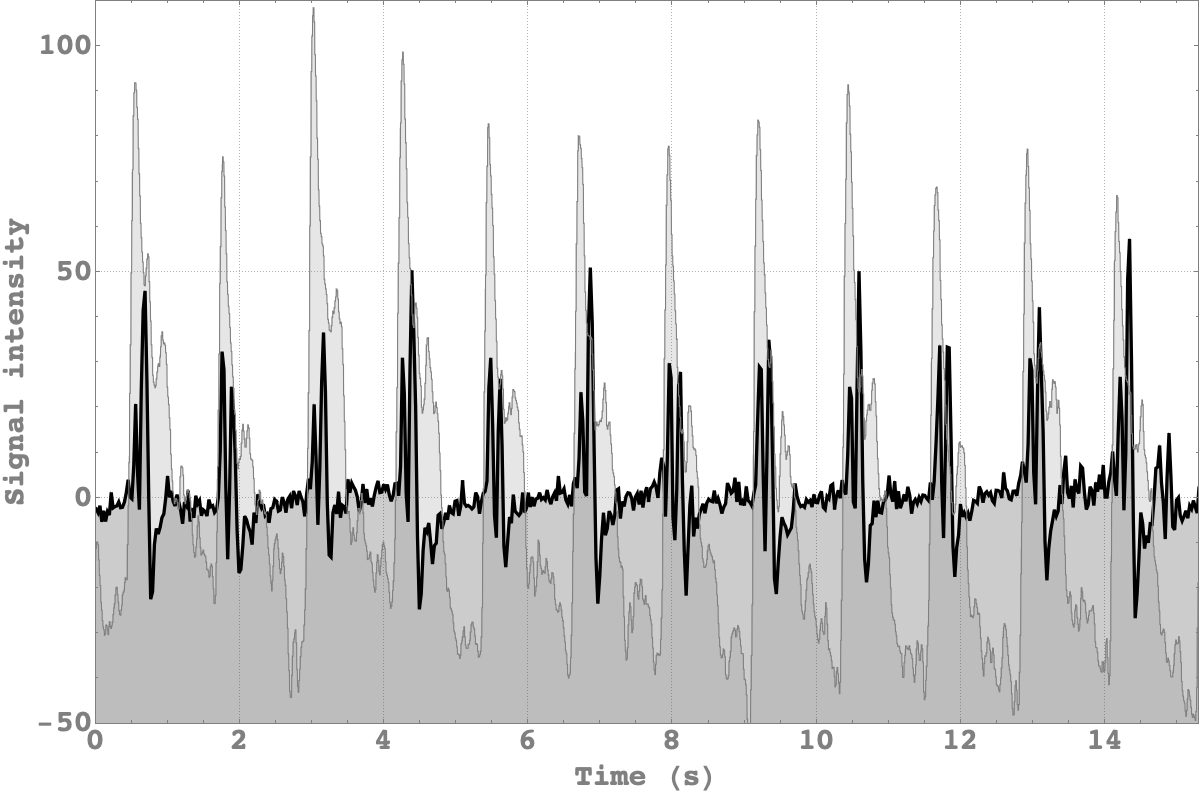}
\caption{{MRI signal time course (Black) during 12 heart cycles compared with
simultaneous oximeter reading of a finger (Grey).
{\label{oximeter}}%
}}
\end{center}
\end{figure} 
\section{Results}\label{sec2}
We used  the EPI time series (as described in section \ref{sec11}) in human volunteers at rest.
The beginning of the sequential  RF-pulses train of the EPI time series  were used to saturate  the magnetisation of the imaging slice. The desired reductions of  the local NMR component were normally reached shortly before  the equilibrium magnetisation. 
\\
Then, we found regular, repeating signal bursts of predominant signal alternations  in single volumes of the brain slices   as shown in Fig.~\ref{oximeter},    where  the   signal peaks of the bursts increased by up to 15 \%. In most cases, the alteration was sequential  from one image acquisition to the next. 
\\
In the following, we will focus  on the NMR contrast mechanism of the signal first, and then how it related to physiology and mind.
\subsection{NMR contrast}  \label{sec21} 
The burst signal had the following properties;
\\
The burst signal alternated  during burst which confirmed that at least two RF pulses were necessary to generate the signal.  The two RF-pulses   always enwrap an asymmetrical gradient interval $G_a T_a$ (Fig. \ref{548938}), which is the basic pulse design to measure   ZQC. \begin{figure}[h!]
\begin{center}
\includegraphics[width=1.00\columnwidth]{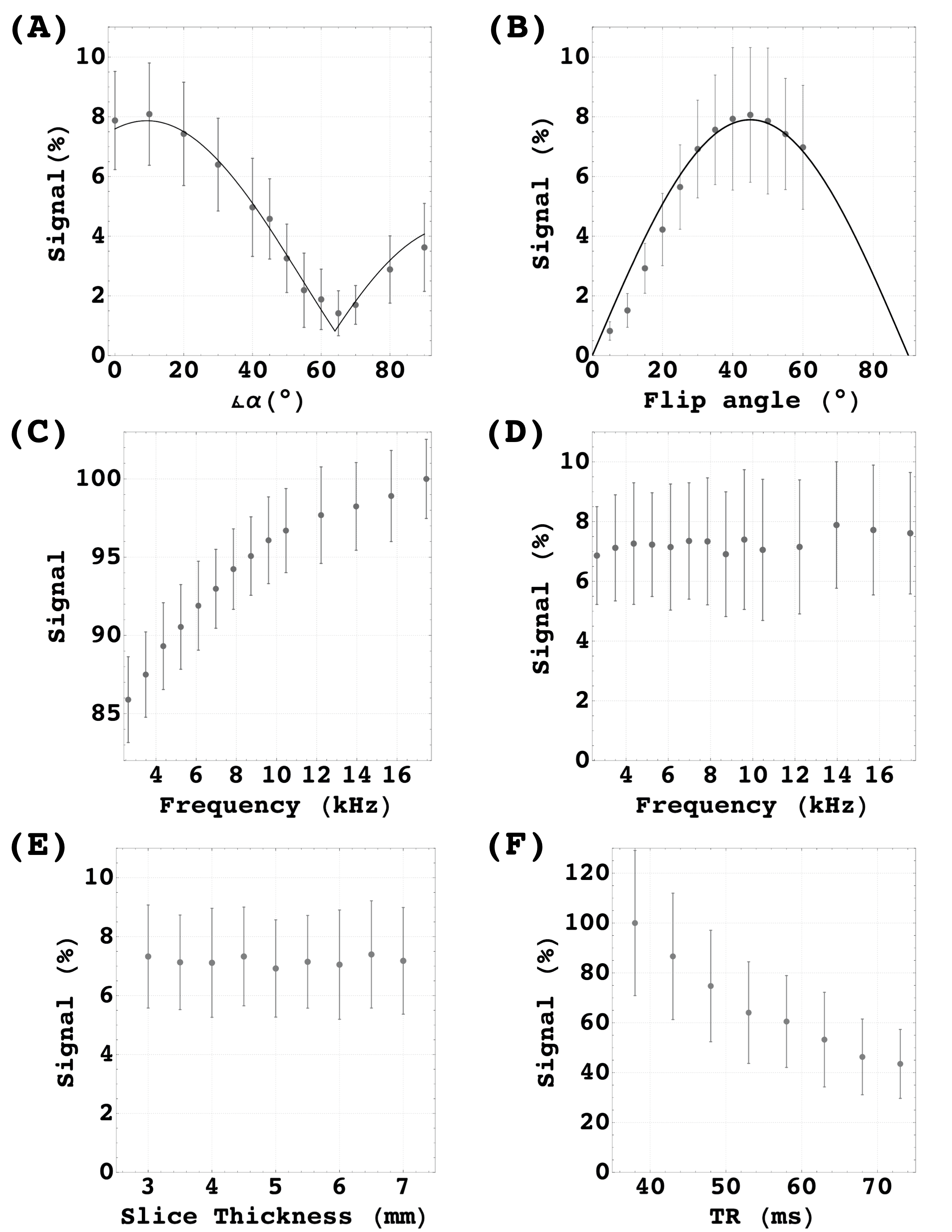}
\caption{{Variation of sequence parameters. Data shows signal averaged over 5
subjects.\textbf{~}Error bars represent the standard deviation from the
mean. \textbf{(A)} Signal intensity plotted against the slice gradient
angulation $\alpha$ in respect to the magnetic main field.~\textbf{(B)} Signal
plotted against flip angle variation. ZQC prediction plotted in Black.
\textbf{(C)} Signal intensity plotted against the frequency offset of
the saturation slices of the BS and~\textbf{(D)}
averaged signal of the AMP.  \textbf{(E)~}Relative signal change plotted against\textbf{~}slice
thickness.~\textbf{(F)~}Signal plotted against repetition time.~
{\label{211018}}%
}}
\end{center}
\end{figure}
The consequential long-range ZQC contrast was  verified  further by altering  sequence parameters. 
\\
For rotating  the asymmetric gradients $G_a$, we found the characteristic angulation dependency  of the dipole-dipole interaction as shown is  (Fig.~\ref{211018}A).
The plot represents the fitted function $\vert(3\cdot cos^2\lbrack\varphi \rbrack-1)\vert $ (adjusted R$^2$ test of goodness-of-fit resulted in R$^2$=0.9958) where $\varphi $ takes the additional gradients in read and phase direction into account. At the magic angle, the burst signals disappeared.    
\\
For the flip angle variation,  we found the predicted signal course  for the ZQC flip angle dependency \unskip~\cite{225062:5277829} which was fitted to the data (R$^2$=0.9964). Predicted maximum  at 45$^\circ$ could be confirmed (Fig~\ref{211018}B).  In contrast, the Ernst-angle \cite{Ernst_1966} which is a good indication for the optimum angle for SQC is around  13$^\circ$ (for T$_1$ = 1.5s). 
\\
For the alteration of the off-resonance frequency, we found    a typical magnetisation transfer contrast (MTC) change for the baseline which depended on the off-resonance frequency (Fig. \ref{211018}C). In contrast, the signal intensity showed the remarkable immunity to MTC as expected for ZQC  \cite{225062:5277825} with
no significant changes in the same frequency range (Fig. \ref{211018}D).
\begin{figure}[h!]
\begin{center}
\includegraphics[width=1.00\columnwidth]{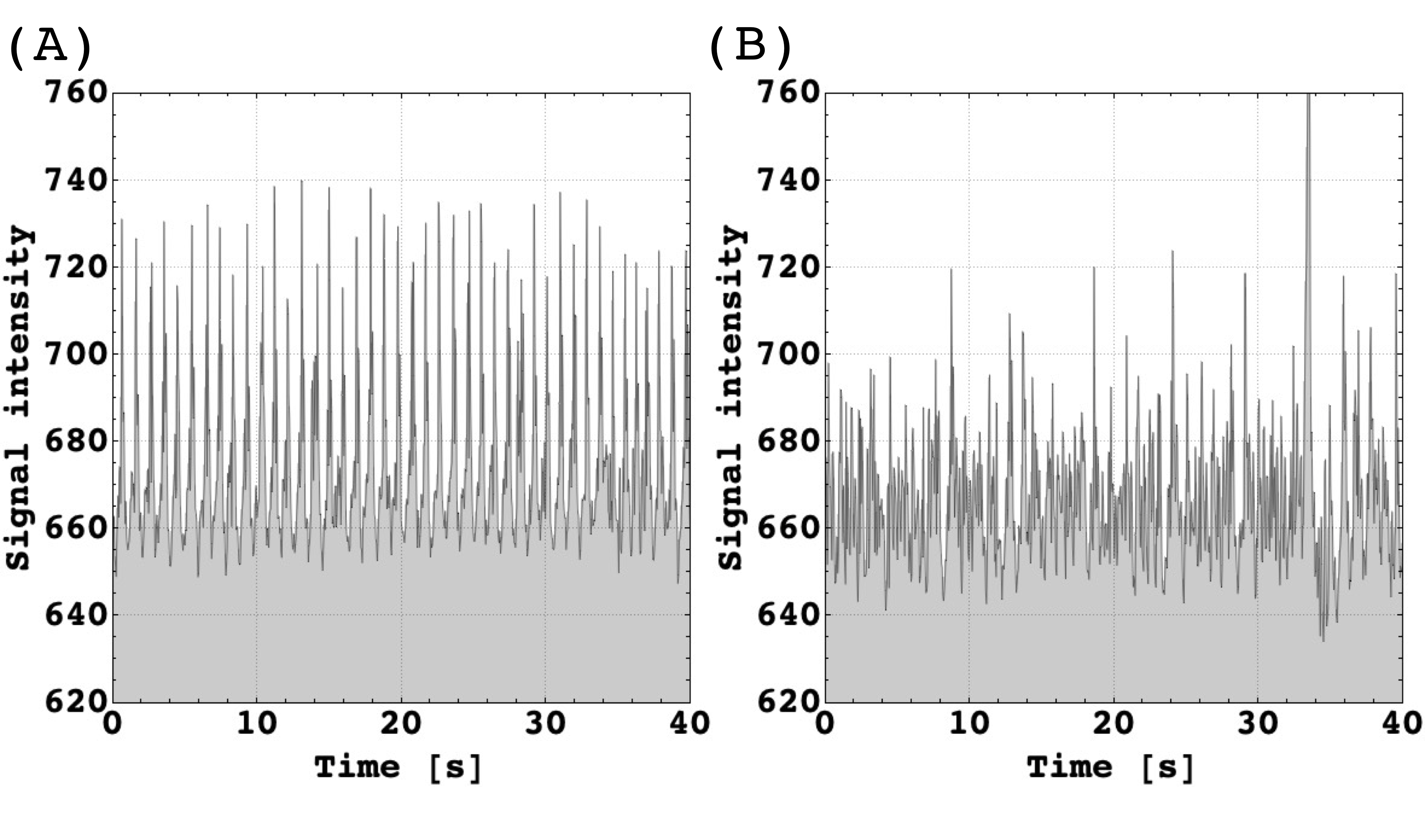}
\caption{{Pattern observed in participant who had reported falling
asleep.~\textbf{(A)} Wake period.~\textbf{(B)} Asleep, ZQC burst signals
declined coincident with an increase of the S/N level. At 34 s, the peak
resulted from short head movement.
{\label{918358}}%
}}
\end{center}
\end{figure}
\\
The effects of the competing effects, the build up of the ZQC on the one hand and de-phasing over time on the other hand, were studied  varying the TR. We found that from 38 ms onwards the signals showed no growth of ZQC. The free induction dominated.
\\
Finally, we varied the slice thickness to study Time-of-flight effects. We found  no significant influence on the relative signal.
\subsection{Physiology and Mind}   
\begin{figure}[h!]
\begin{center}
\includegraphics[width=0.94\columnwidth]{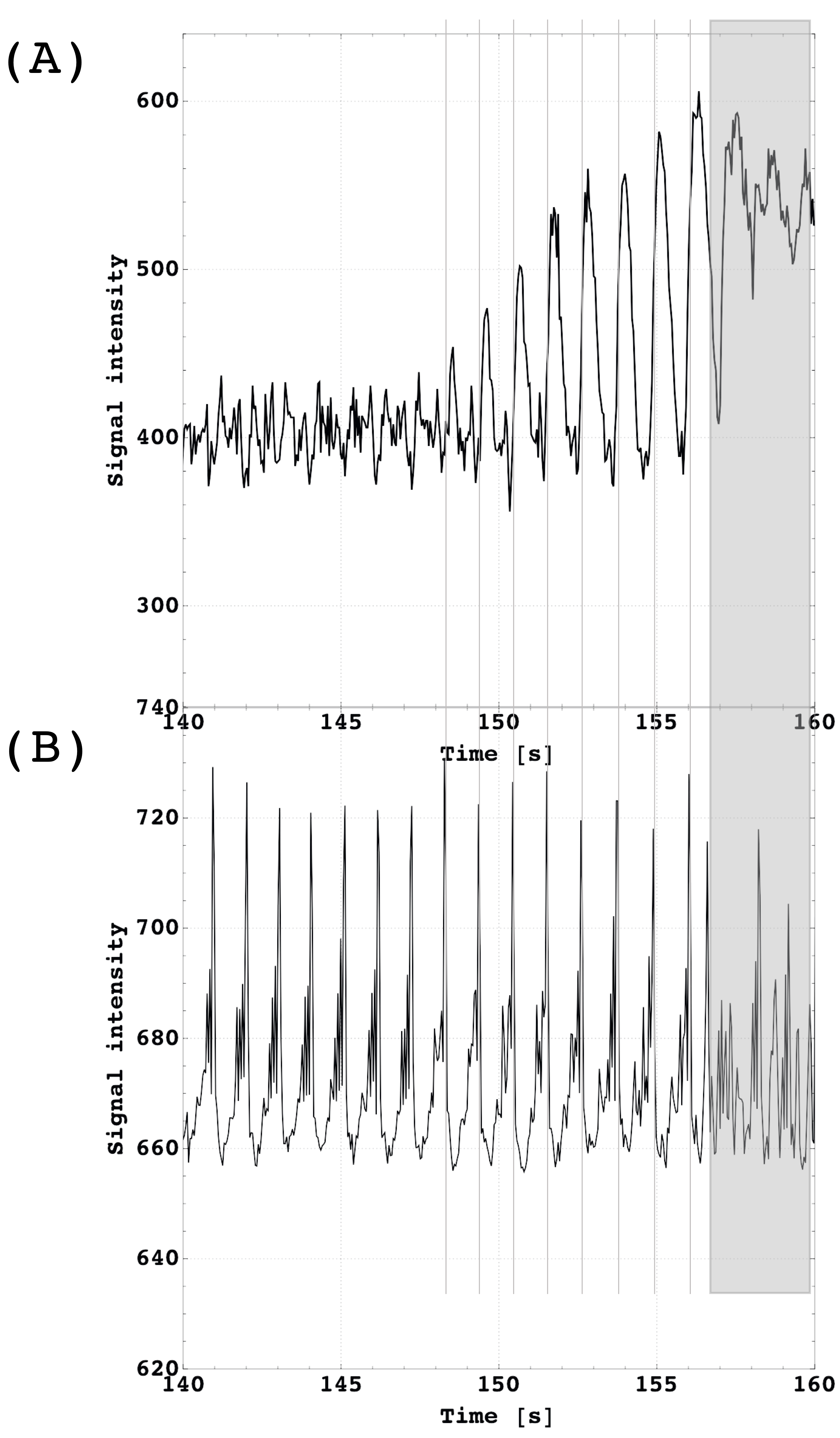}
\caption{{\textbf{(A)} Signal time course of an imaging voxel located next to the
superior sagittal sinus demonstrates the blood flow increase in response
to the CO\textsubscript{2} challenge (breath-holding). In contrast to
the vein signal, the corresponding ZQC signals~\textbf{(B)} showed no
response to CO\textsubscript{2} activity. Breath-holding started at
140s. Volunteers were instructed to reduce any movement as long as
possible (here until at 157s). From 157s, the signal breakdown was
subject to movement.~
{\label{372115}}%
}}
\end{center}
\end{figure}

\begin{figure}[h!]
\begin{center}
\includegraphics[width=1.00\columnwidth]{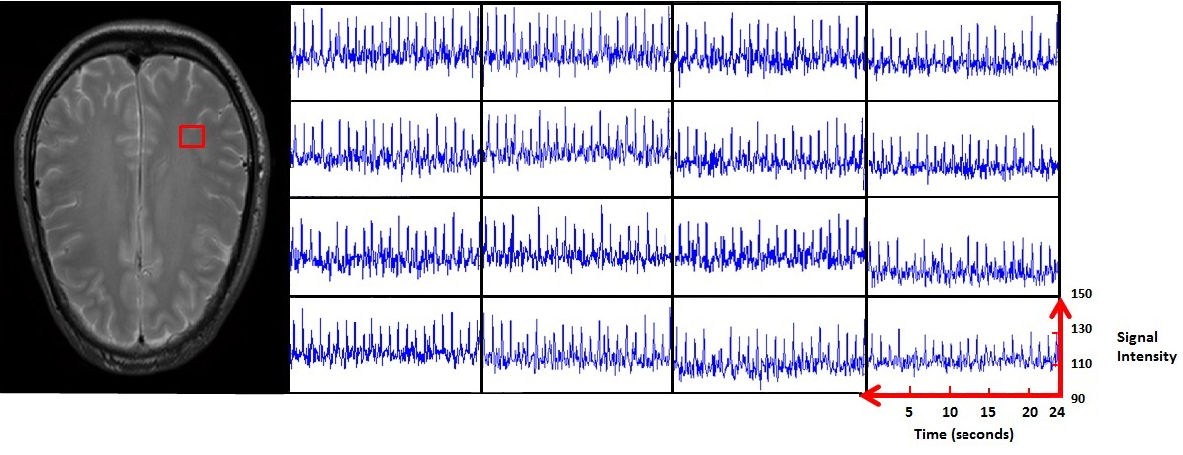}
\caption{{4x4 voxel matrix randomly picked. On the left,~the red square shows
location in the brain slice. On the right, 16 corresponding signal time
courses displaying the local tissue responses over a time period of 24s.
{\label{972921}}%
}}
\end{center}
\end{figure}

\begin{figure}[h!]
\begin{center}
\includegraphics[width=1.00\columnwidth]{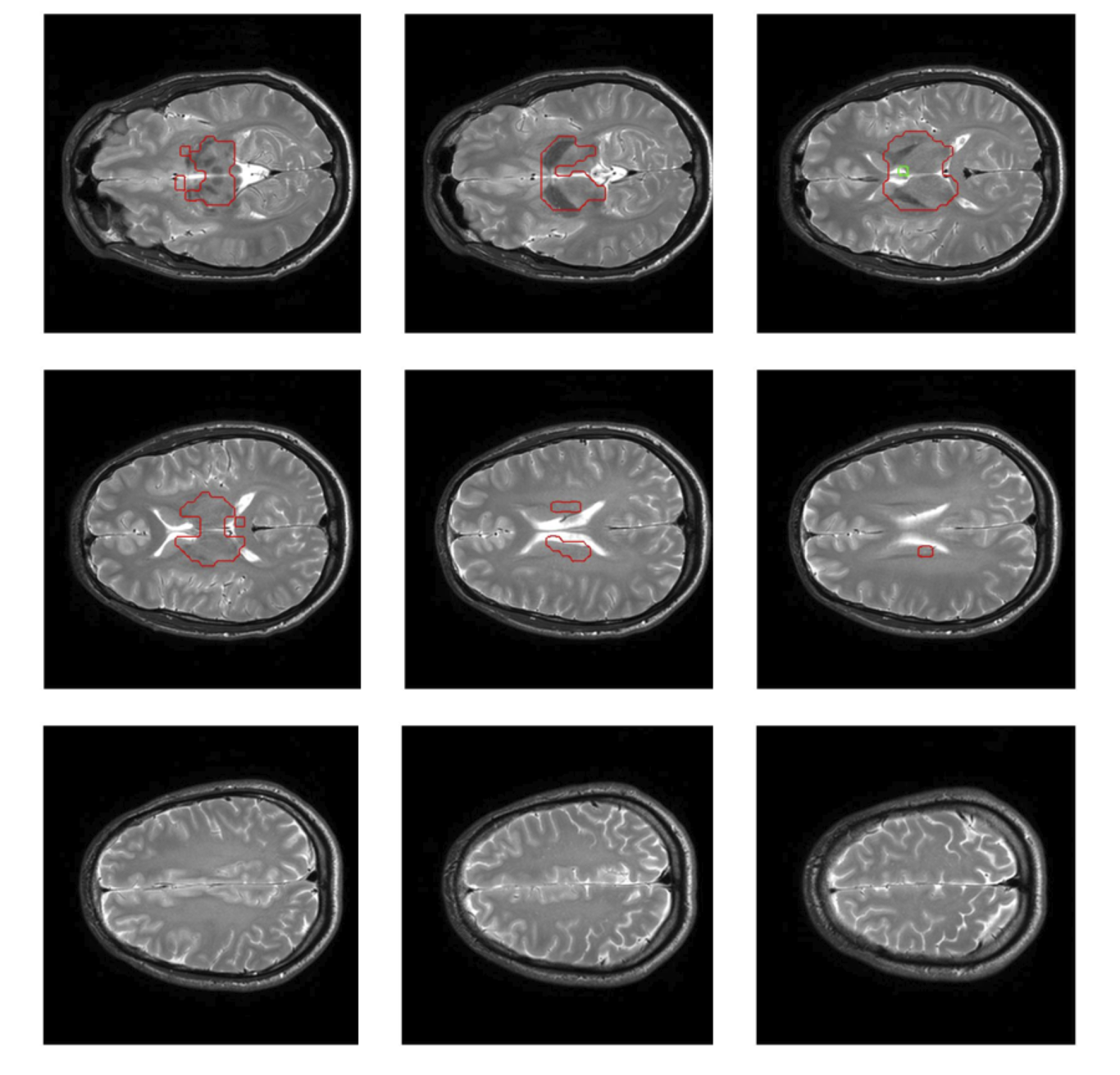}
\caption{{9 Anatomical slices which correspond to the positioning of the EPI time
series. Tissue surrounded by red drawing showed no ZQC bursts.
{\label{640443}}%
}}
\end{center}
\end{figure}

\begin{figure}[h!]
\begin{center}
\includegraphics[width=.700\columnwidth]{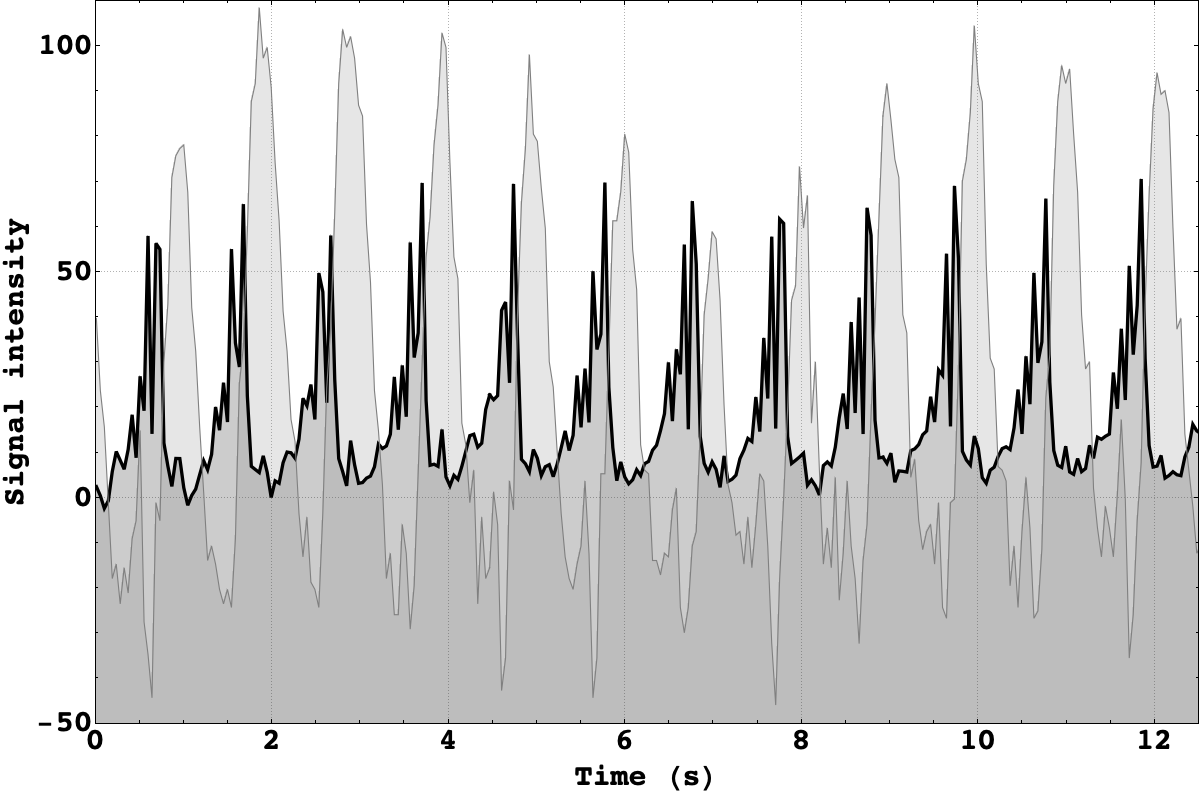}
\caption{{Whole-slice averaged signal time course (black line) which was selected
by a mask over 12 heart cycles. Signal of the Superior sagittal sinus
(grey line) as reference time frame demonstrates the instant breakdown
of quantum coherence with the beginning outflow.~
{\label{125409}}%
}}
\end{center}
\end{figure}

\begin{figure}[h!]
\begin{center}
\includegraphics[width=1.00\columnwidth]{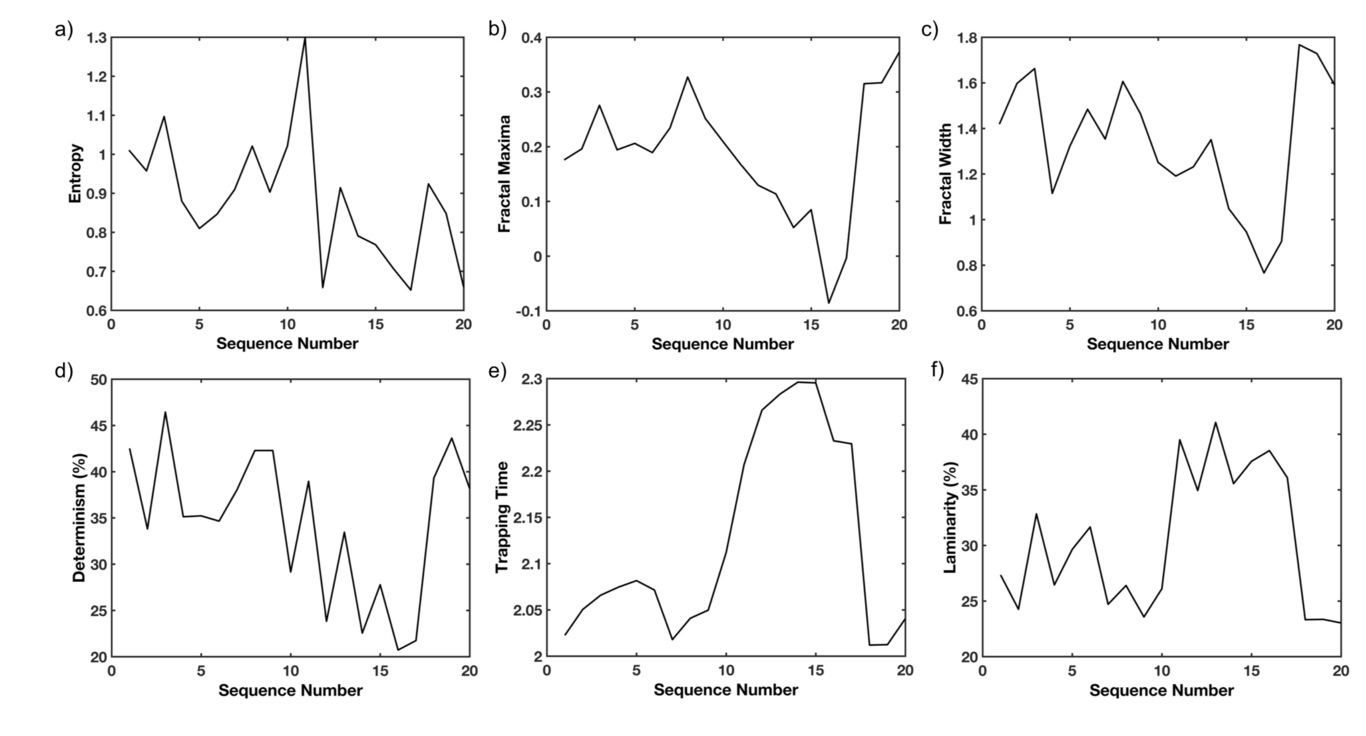}
\caption{{Case study: Results of a Recurrence Quantication Analysis and a
Multifractal Detrended Fluctuation Analysis using 20 time periods a 45s
over a total time period of 21 minutes. \textbf{\textbf{(}a)} Entropy
(Ent) is computed as the Shannon entropy of the distribution of the
repeating pattern of the system. If a signal has high entropy it
exhibits diversity in short and long duration
periodicities.~\textbf{(b-c)}~The multifractal spectrum identifies the
deviations in fractal structure within time periods with large and small
fluctuations.~\textbf{(d)}~Determinism (DET) represents a measure that
quantifies repeating patterns in a system and it is a measure of its
predictability. Regular, periodic signals, such as sine waves, will have
higher DET values, while uncorrelated time series will cause low
DET.~\textbf{(e)} Trapping Time (TT) represents the average time the
system remains in a given state and it is a measure of the stability of
the system.~\textbf{(f)} Laminarity (Lam) determines the frequency of
transitions from one state to another, without describing the length of
these transition phases. It indexes the general level of persistence in
some particular state of one of the time-series.~
{\label{642362}}%
}}
\end{center}
\end{figure}

The periods of signal bursts  repeated with the same rate as the heart-beat. We used three temporal reference systems;  (a) a finger pulse oximetry, (b) an electrocardiogram (ECG), and (c) the time-of-flight signal of a voxel placed in the superior sagittal sinus. The signal bursts  appeared  with the  pulse from the finger pulse oximetry (Fig.~\ref{oximeter}). In relation to the ECG, we found using the Cross-Recurrence Quantification Analysis that the  maximum burst signal was delayed by 0.3s on average. With the start of the venous outflow, the bursts always ended as shown in Fig. \ref{372115} and Fig. \ref{125409}. 
\\
Regarding the  duration of the bursts under normal conditions, we mostly observed two sequential  peaks which equaled  4 TRs  adding up to a time period of 180ms. We also saw longer periods building up to 10 TRs (see Fig. \ref{372115}B) extending the period  to  450ms.\\
We located  the bursts in brain tissue of all slices  except around the periventricular area (probably due to movement induced by  ventricular pulsation  in those  regions \unskip~\cite{225062:5067666}) as illustrated  in Fig. \ref{640443}. \\
The global aspect  conformed with another interesting feature; the  signal could be restored while being averaged over the entire tissue component of the imaging slice (Fig.~\ref{oximeter} and Fig.~\ref{372115}B, single voxel time course are shown in Fig. \ref{972921}).\\
We also found that the  signal   did not respond to the CO$_{2}$ challenge (Fig.~\ref{372115}B) in contrast to the SQC  signal 
from the voxel including the superior sagittal sinus (Fig. \ref{372115}A) which indicated the   blood flow response.
\\
During our studies, we also realised that the signal  depended on awareness and awakening.
In seven participants from whom two had reported to have fallen asleep, we found that the  signal pattern declined  as shown in Fig.~\ref{918358}.  For the final data acquisition, all participants had been asked to stay awake during the imaging protocol. At this point, we no longer detected a sleep pattern. 
\\
In a case study, we observed the   pattern change over a period of 20 minutes which showed a  gradual transition from awake to asleep as shown in the appendix  at Fig.~\ref{sleep6}.  \\
We used   Recurrence Quantification Analysis and  Multifractal Detrended Fluctuation Analysis to   illustrate the difference between wakefulness and the slow decline during the falling asleep period. The analysis shows that periodicity, large and small fluctuations, repeating patterns and their predictability, and the stability of the system were changing over the observation period (Fig.~\ref{642362}). 
\section{Methods}\label{sec11}
\begin{figure}[h]
\begin{center}
\includegraphics[width=1.00\columnwidth]{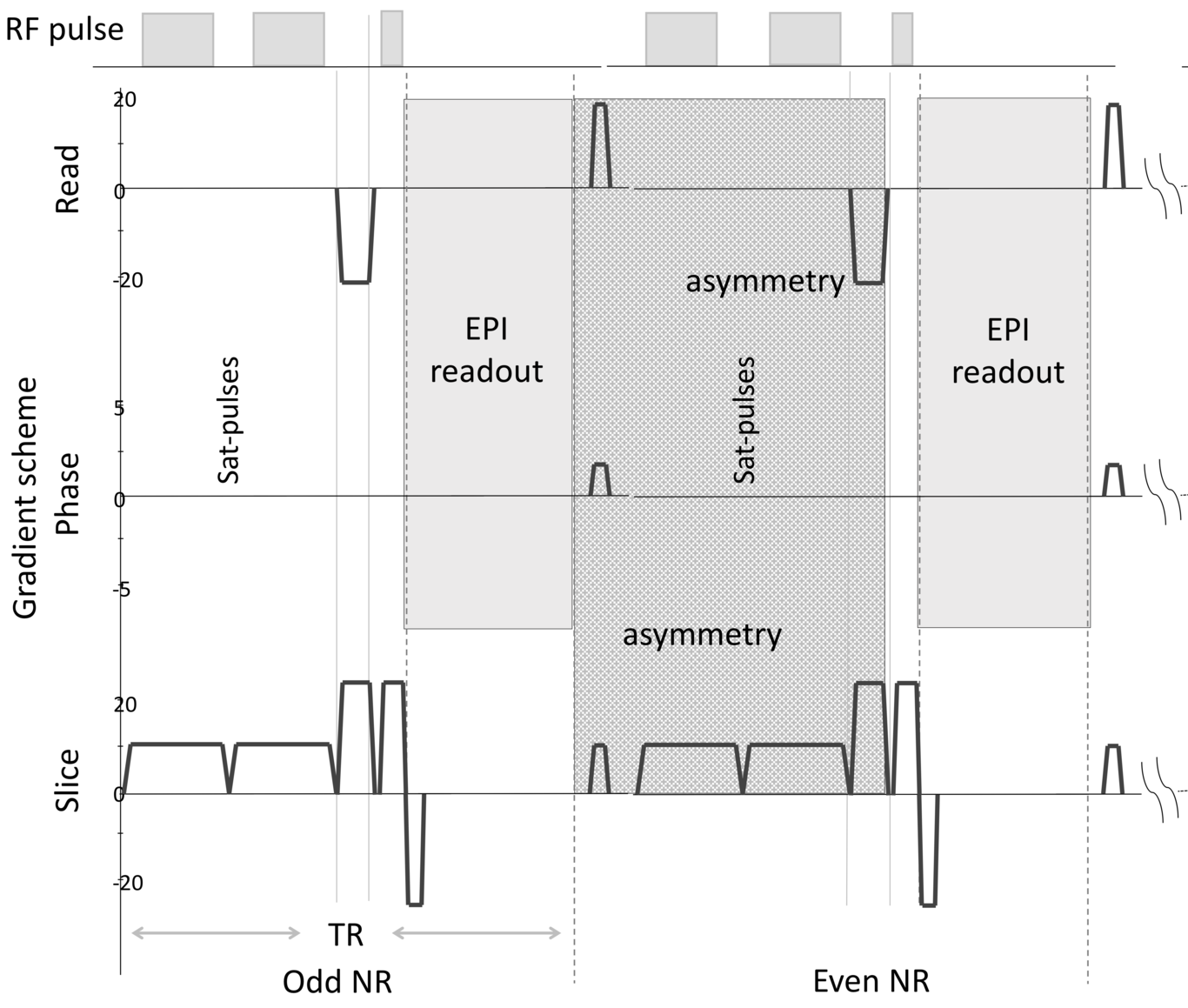}
\caption{{Radio frequency (RF) and Gradient scheme of two consecutive EPI
acquisition. The ``asymmetry'' field includes all asymmetric gradients involved in the ZSE.
}}
\label{548938}
\end{center}
\end{figure}
We studied 40 subjects (between 18 and 46 years old) using a 3 Tesla  whole-body MRI scanner (Philips, The Netherlands) which was operated with a 32-channel array receiver coil. 
\\
Imaging protocols  were approved by Trinity College Dublin School of Medicine Research Ethics Committee. 
\\
All participants of final data acquisition were asked to stay awake and stay still during the imaging protocol, or to report any failure to do so.
\\
Fast  gradient-echo EPI (GE-EPI) time series  were carried out which had been optimised over a wide range of participants. The finalised parameters were as follows: FA = 45$^\circ$, TR = 45 ms, TE = 5, voxel size = 3.5 x 3.5 x 3.5 mm, matrix size = 64x64, SENSE factor = 3, bandwidth readout direction = 2148 Hz, saturation pulse thickness/distance = 5/20mm. 
\\
Two saturation pulses   placed parallel to the imaging slice (Fig.~\ref{548938}) were added which allowed us to  vary  long-range correlation of the ZSE and MTC.
Saturation gradients had a time integral (length x strength) of GT$_{s} = $ 5.1 ms x 6.25 mT/m, the crusher gradients in read and slice direction of  GT$_{c} = $ 1.3 ms x 25 mT/m, the slice rephase gradient of GT$_{r} = $ 0.65 ms x 25 mT/m, and the slice termination gradient of GT$_{t} = $ 0.65 ms x 15 mT/m. Gradients timing and arrangements are shown in Fig. \ref{548938}. Gradients relevant for ZSE are shown in the asymmetry field and are marked with indices t, c, r, and s for identification. We rotated  the asymmetric gradients in respect to the magnet field starting from coronal 0$^\circ$ to axial 90$^\circ $in twelve  steps; slice angulation $\alpha$ related to the angulation from the  spin-spin interaction as \\  $\varphi=\alpha-\tan^{-1} \left([\textnormal{GT}_{c}-\textnormal{GT}_{r}]/[2 \cdot\textnormal{GT}_{s}+\textnormal{GT}_{c}+\textnormal{GT}_{t}]\right)=\alpha-\textnormal{9.6}^\circ$.  Further, we varied the correlation distance   via altering the amplitude and the duration of  the saturation gradients.\\
We also altered the following sequence parameters in  pseudo-randomised orders: \\
 (a) variation of the flip angle from 5$^\circ $ to 60$^\circ $ in steps of 5$^\circ $(60$^\circ$ was the power limit by the specific absorption rate (SAR)).\\
 (b) the off-resonance frequency  was varied as [2.62, 3.49, 4.36, 5.23, 6.11, 6.98, 7.84, 8.73, 9.60, 10.47, 12.22, 13.96, 15.71, 17.45] kHz.\\
 (c) slice thickness from 3 mm to 7 mm in steps of 0.5 mm.\\
 (d) repetition time (TR) varied from 38 ms to 73 ms in steps of 5 ms.\\
 Further, we explored the signal distribution  over the entire brain. 
 9 slices (in 5 volunteers) were acquired at different positions, with slices   from bottom to the top covering all anatomical regions. \\
In a breath-holding challenge,  four participants  were asked to stop breathing for 20 s without taking a deep breath. Body movements were reduced through  multiple cushions immobilizing the head.\\
For the time reference analysis, we used Cross-Recurrence Quantification Analysis \cite{MARWAN2002299} to calculate the delay between the R-wave in electrocardiogram (ECG) and the MRI signal. For the calculation, we used  the CRP Toolbox  \cite{webpage,MARWAN2007237} for Matlab \cite{MATLAB:2014}. 
\\
For the NMR contrast analysis, we used the averaged maximum peak of the burst and the signals between   bursts as baselines. Calculations were performed using the routine by Gomes et al. \unskip~\cite{225062:5042988} which was implemented in Matlab \cite{MATLAB:2014}. Preprocessing included the following; Rescaling, which was applied to all data sets before any analysis using the MR vendor's instructions. Visual inspection of  average time series  in search for irregularities which were manually removed from the analysis leaving the rest of the time series unaltered. Manual segmentation was used to create a mask to remove cerebral spinal fluid (CSF) contributions. The first 100  of 1000 scans were removed to avoid signal saturation effects. The manual segmentation of the masks was eroded to avoid partial volume effects at the edges. 
\\
For the analysis of sleeping pattern, we used a  Recurrence Quantification Analysis and a Multifractal Detrended Fluctuation Analysis (for detailed description see Lopez-Perez et al.~\cite{Perez:2020aa}).
\\  
All data graphics were created with Mathematica \cite{Mathematica}. Data and source code for analysis are available at www.github.com/Mirandeitor/Entanglement-witnessed-in-the-human-brain.
\section{Discussion}\label{sec12}
The aim of this study was to find evidence that brain functions can create entanglement in auxiliary quantum systems. 
\\  
Thereby, we employed a hybrid  MRI sequence which could contain SQC and ZQC, simultaneously. We found that the heart pulsation evoked NMR signal burst with every heartbeat. We were able to show in section \ref{sec21}  that the signal contrast originated from spin-spin interactions. Therefore, it is possible that we witnessed quantum entanglement. 
\\ 
However, NMR signals can be  altered by many physiological changes. 
Ultimately, we had to prove that the signal bursts was not a “classical” ZQC. 
\\
As mentioned above, classical ZQC have corresponding contrasts  in SQC, namely  T2* relaxation and diffusion. Both contrasts alter during the heart cycle. 
\\
However, T2* changes have  shown a different temporal (shifted by more than half of the cycle time in respect to the ZQC signal) and spatial response (higher signal at blood vessel)  \cite{DAGLI1999407}. The tissue response at around 2$\%$ is much lower than during functional activation. In contrast, functional activations showed no significant changes in the ZQC  burst signal and only minimal signal increases at the baseline \cite{davidprivate}.  Therefore, we can conclude that local field-inhomogeneity changes are ineligible as a signal source. \\
Besides local field-inhomogeneity, ZQC depends on   order \cite{225062:5187628} and rotational symmetries \cite{225062:5067412,225062:5041128} which can be probed with diffusion MRI. The order may correlate with the apparent diffusion coefficient (ADC), while the fractional anisotropy (FA) indicates the rotational symmetry breaking. In praxis, MQC signals are higher at decreased ADC and increased FA.
\\
Nakamura et al. \cite{Nakamura:2009uq} have  shown that  the temporal changes of the  ADC-values are in phase with the  intracranial volume change while FA-values show a shift by 180$^\circ$. Our ZQC signals coincided with the transition phase from the highest to the lowest ADC (and vice versa for the FA). From those results, we can conclude that the theoretical optimum (ADC minimal, FA maximal) for a classical ZQC is outside  the time window of the ZQC bursts. \\
We conclude, that our observation has no corresponding SQC contrast. 
\\
Further, the signals surpassed the classical bound by far. For “classical fluids”, the S/N of ZQC compared to the conventional MRI signal (SQC)  only reaches up to 0.05 at 4 Tesla, experimentally \cite{225062:5289250,Rizi_2000}. Our sequence was suboptimal because we replaced  a 90$^\circ$  by a 45$^\circ$ RF-pulse (reduction by factor 2), used a 3 Tesla field, and the evolution time was shorter.  
Therefore, we can conclude that in combination with  the EPI readout, that classical ZQC signals weren't detectable with our sequence. Even more, in the above argument, we discussed baseline signals. Our observations showed fluctuations which, if translated to a classical ZQC, would then be serval magnitudes higher than the actual baseline ZQC signal.  
\\
Although, we found that the evoked bursts disappeared at the magic angle which means they have no SQC component, cardiac pulsation can cause flow and motion  effects which we further investigated.
\\
We varied slice thickness and TR as possible sequence parameters, which are sensitive to time-of-flight effects. For the slice thickness,  the relative signal did not vary significantly (Fig.~\ref{211018}E), for   the repetition time, we found the free induction decay  dominating the  decline (Fig.~\ref{211018}F). Furthermore, when we varied the blood flow with the help of a CO$_2$-challenge (Fig.~\ref{372115}), we found no significant response of the burst signal amplitude. 
\\
Further, the fact the signal bursts have no significant  SQC component (Fig ~\ref{211018}A at the magic angle), we can in principle exclude all SQC contrast mechanism including changes in T$_1$ and T$_2$ relaxation, line narrowing, or magnetic field shifts. 
\\
Above, we  have established  that conventional MR sequences, be it SQC or MQC, are unable to detect the observed signal bursts. Further, we found that the signal amplitude is above the bound which could classically be reached. 
\\
By now, it is clear that the evoked signals can only be observed if the necessary condition, that the magnetisation is highly saturated, is met. 
\\
We also considered what we called the sufficient condition above. We found that the timing of the signal bursts coincided the first cluster of the HEP \cite{KERN2013178}. Like the timing, the signal intensity also showed a similar dependence to conscious awareness in this time window \cite{10.1093/sleep/zsab100,PARK2019502}.
\\
In another study,  L\'{o}pez P\'{e}rez et al. \cite{Perez:2020aa} have  shown that the complexity of burst signals correlate with psychological test results in short-term memory. This relation is also known in HEPs.
\\
To our knowledge, both, the direct correlation to conscious awareness and short-term memory, are unreported in classical MRI experiments. It underpins that our findings are from the same origin as HEPs and that there is no classical correlate in MRI.  
\section{Conclusion}\label{sec13}
The aim of this study was to show that the brain is non-classical. 
We assumed that  unknown brain functions exist which can mediate  
entanglement between auxiliary quantum systems. The experimental detection of such an entanglement created by the brain would then be sufficient to prove cerebral non-classicality.
\\
We found experimental evidence that such entanglement  creation occurs  as part of  physiological and cognitive processes.
\\
We argued that the ZQC signals are non-local because (a) ZQC signals are above the classical bound, and (b) the signals have no SQC and MQC\footnote{using  the conventional MQC sequence design} correlates. Further, we could confirm that the signals are only detectable in combination with reduced classical signals (necessary condition), and that they resemble HEPs which are below verifiability in conventional MRI (sufficient condition).
\\
Our findings disapprove the unproven statement that quantum entanglement or  coherence can't survive in the hot and wet environment of the brain.
\\
Beyond the fundamental question we tried to answer here, we found  an undiscovered NMR contrast, which can detect brain activity beyond conventional functional MRI. It may have  interesting applications in psychology and  medicine.

\section*{Declarations}

\begin{itemize}
\item Funding\\
This  research project was funded by Science Foundation Ireland from 2011-15 (SFI-11/RFP.1/NES/3051) and supported by Trinity College Institute of Neuroscience.
\item Authors' contributions\\
Christian Matthias Kerskens: conceptualisation, methodology (physics), writing (original draft), supervision, funding acquisition. \\
David López Pérez: methodology (analysis), software, acquisition of data, data curation.
\end{itemize}

\pagebreak
\newpage
\pagebreak

\section{Extended Data}\label{secA1}
\begin{figure}[h!]
\begin{center}
\includegraphics[width=1.00\columnwidth]{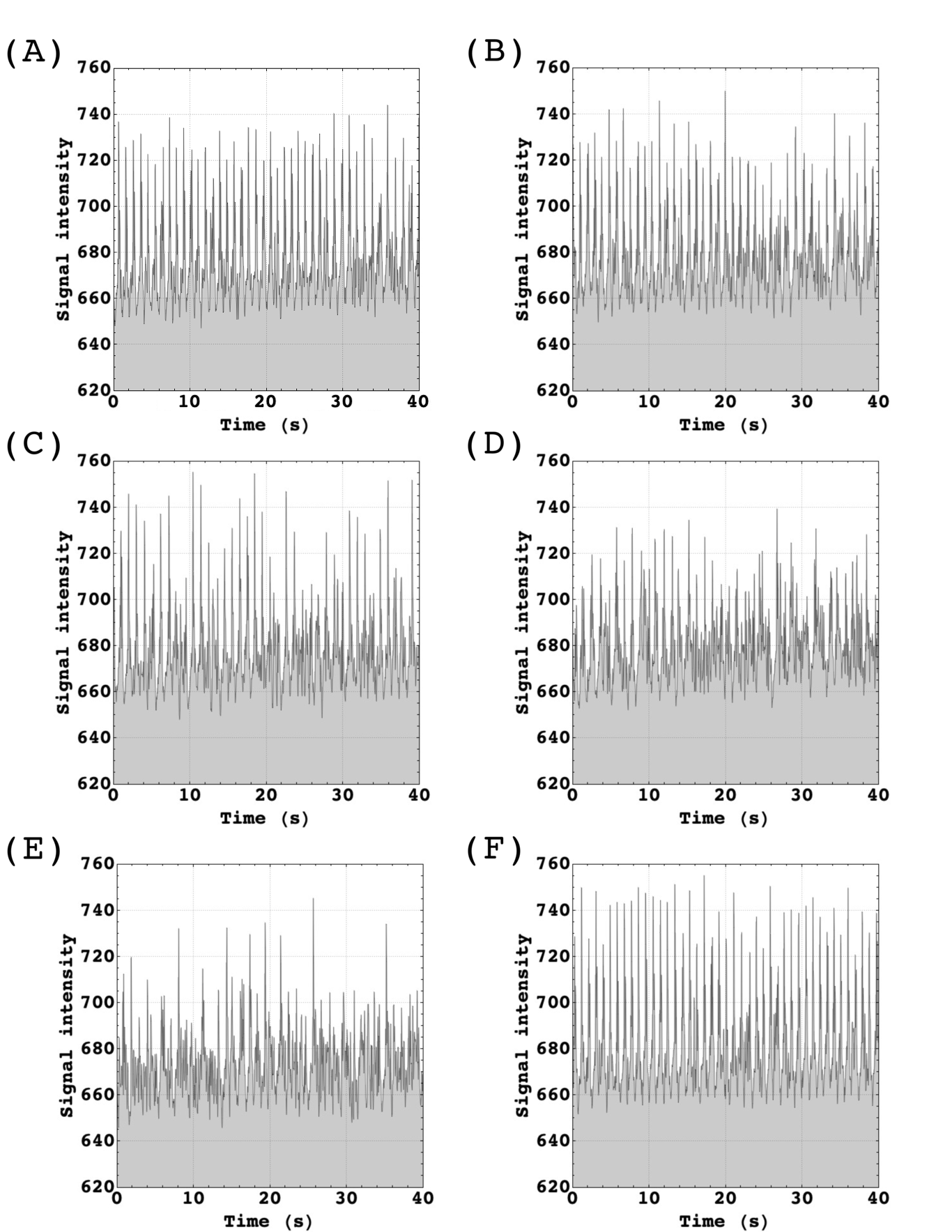}
\caption{{Case~ study: ZQC burst pattern observed in participant who had reported
falling asleep. Starting point of time series at~\textbf{(A)}
16:26:29~\textbf{(B)} 16:29:47~\textbf{(C)~}16:30:54 \textbf{(D)}
16:34:13~\textbf{(E)} 16:37:32~\textbf{(F)} 16:40:49 (awake, subject
communicated with radiographer before scan).%
}}
\label{sleep6}
\end{center} 
\end{figure}




\selectlanguage{english}
\clearpage
\bibliographystyle{apsrev4-1}
\bibliography{/Users/christian/Documents/lib/lib}

\begin{thebibliography}{45}%
\makeatletter
\providecommand \@ifxundefined [1]{%
 \@ifx{#1\undefined}
}%
\providecommand \@ifnum [1]{%
 \ifnum #1\expandafter \@firstoftwo
 \else \expandafter \@secondoftwo
 \fi
}%
\providecommand \@ifx [1]{%
 \ifx #1\expandafter \@firstoftwo
 \else \expandafter \@secondoftwo
 \fi
}%
\providecommand \natexlab [1]{#1}%
\providecommand \enquote  [1]{``#1''}%
\providecommand \bibnamefont  [1]{#1}%
\providecommand \bibfnamefont [1]{#1}%
\providecommand \citenamefont [1]{#1}%
\providecommand \href@noop [0]{\@secondoftwo}%
\providecommand \href [0]{\begingroup \@sanitize@url \@href}%
\providecommand \@href[1]{\@@startlink{#1}\@@href}%
\providecommand \@@href[1]{\endgroup#1\@@endlink}%
\providecommand \@sanitize@url [0]{\catcode `\\12\catcode `\$12\catcode
  `\&12\catcode `\#12\catcode `\^12\catcode `\_12\catcode `\%12\relax}%
\providecommand \@@startlink[1]{}%
\providecommand \@@endlink[0]{}%
\providecommand \url  [0]{\begingroup\@sanitize@url \@url }%
\providecommand \@url [1]{\endgroup\@href {#1}{\urlprefix }}%
\providecommand \urlprefix  [0]{URL }%
\providecommand \Eprint [0]{\href }%
\providecommand \doibase [0]{http://dx.doi.org/}%
\providecommand \selectlanguage [0]{\@gobble}%
\providecommand \bibinfo  [0]{\@secondoftwo}%
\providecommand \bibfield  [0]{\@secondoftwo}%
\providecommand \translation [1]{[#1]}%
\providecommand \BibitemOpen [0]{}%
\providecommand \bibitemStop [0]{}%
\providecommand \bibitemNoStop [0]{.\EOS\space}%
\providecommand \EOS [0]{\spacefactor3000\relax}%
\providecommand \BibitemShut  [1]{\csname bibitem#1\endcsname}%
\let\auto@bib@innerbib\@empty
\bibitem [{\citenamefont {Schoenlein}\ \emph {et~al.}(1991)\citenamefont
  {Schoenlein}, \citenamefont {Peteanu}, \citenamefont {Mathies},\ and\
  \citenamefont {Shank}}]{doi:10.1126/science.1925597}%
  \BibitemOpen
  \bibfield  {author} {\bibinfo {author} {\bibfnamefont {R.~W.}\ \bibnamefont
  {Schoenlein}}, \bibinfo {author} {\bibfnamefont {L.~A.}\ \bibnamefont
  {Peteanu}}, \bibinfo {author} {\bibfnamefont {R.~A.}\ \bibnamefont
  {Mathies}}, \ and\ \bibinfo {author} {\bibfnamefont {C.~V.}\ \bibnamefont
  {Shank}},\ }\href {\doibase 10.1126/science.1925597} {\bibfield  {journal}
  {\bibinfo  {journal} {Science}\ }\textbf {\bibinfo {volume} {254}},\ \bibinfo
  {pages} {412} (\bibinfo {year} {1991})},\ \Eprint
  {http://arxiv.org/abs/https://www.science.org/doi/pdf/10.1126/science.1925597}
  {https://www.science.org/doi/pdf/10.1126/science.1925597} \BibitemShut
  {NoStop}%
\bibitem [{\citenamefont {Keller}\ and\ \citenamefont
  {Vosshall}(2004)}]{Keller:2004uh}%
  \BibitemOpen
  \bibfield  {author} {\bibinfo {author} {\bibfnamefont {A.}~\bibnamefont
  {Keller}}\ and\ \bibinfo {author} {\bibfnamefont {L.~B.}\ \bibnamefont
  {Vosshall}},\ }\href {\doibase 10.1038/nn1215} {\bibfield  {journal}
  {\bibinfo  {journal} {Nature Neuroscience}\ }\textbf {\bibinfo {volume}
  {7}},\ \bibinfo {pages} {337} (\bibinfo {year} {2004})}\BibitemShut {NoStop}%
\bibitem [{\citenamefont {Hore}\ and\ \citenamefont
  {Mouritsen}(2016)}]{doi:10.1146/annurev-biophys-032116-094545}%
  \BibitemOpen
  \bibfield  {author} {\bibinfo {author} {\bibfnamefont {P.~J.}\ \bibnamefont
  {Hore}}\ and\ \bibinfo {author} {\bibfnamefont {H.}~\bibnamefont
  {Mouritsen}},\ }\href {\doibase 10.1146/annurev-biophys-032116-094545}
  {\bibfield  {journal} {\bibinfo  {journal} {Annual Review of Biophysics}\
  }\textbf {\bibinfo {volume} {45}},\ \bibinfo {pages} {299} (\bibinfo {year}
  {2016})},\ \bibinfo {note} {pMID: 27216936},\ \Eprint
  {http://arxiv.org/abs/https://doi.org/10.1146/annurev-biophys-032116-094545}
  {https://doi.org/10.1146/annurev-biophys-032116-094545} \BibitemShut
  {NoStop}%
\bibitem [{\citenamefont {Wiltschko}(1968)}]{wiltschko1968}%
  \BibitemOpen
  \bibfield  {author} {\bibinfo {author} {\bibfnamefont {W.}~\bibnamefont
  {Wiltschko}},\ }\href@noop {} {\bibfield  {journal} {\bibinfo  {journal} {Z
  Tierpsychol}\ }\textbf {\bibinfo {volume} {25}},\ \bibinfo {pages} {537}
  (\bibinfo {year} {1968})}\BibitemShut {NoStop}%
\bibitem [{\citenamefont {Sechzer}\ \emph {et~al.}(1986)\citenamefont
  {Sechzer}, \citenamefont {Lieberman}, \citenamefont {Alexander},
  \citenamefont {Weidman},\ and\ \citenamefont {Stokes}}]{SECHZER19861258}%
  \BibitemOpen
  \bibfield  {author} {\bibinfo {author} {\bibfnamefont {J.~A.}\ \bibnamefont
  {Sechzer}}, \bibinfo {author} {\bibfnamefont {K.~W.}\ \bibnamefont
  {Lieberman}}, \bibinfo {author} {\bibfnamefont {G.~J.}\ \bibnamefont
  {Alexander}}, \bibinfo {author} {\bibfnamefont {D.}~\bibnamefont {Weidman}},
  \ and\ \bibinfo {author} {\bibfnamefont {P.~E.}\ \bibnamefont {Stokes}},\
  }\href {\doibase https://doi.org/10.1016/0006-3223(86)90308-2} {\bibfield
  {journal} {\bibinfo  {journal} {Biological Psychiatry}\ }\textbf {\bibinfo
  {volume} {21}},\ \bibinfo {pages} {1258} (\bibinfo {year}
  {1986})}\BibitemShut {NoStop}%
\bibitem [{\citenamefont {Li}\ \emph {et~al.}(2018)\citenamefont {Li},
  \citenamefont {Lu}, \citenamefont {Yang}, \citenamefont {Tao}, \citenamefont
  {Xu}, \citenamefont {Wang}, \citenamefont {Fu}, \citenamefont {Liu},
  \citenamefont {Chummum},\ and\ \citenamefont {Zhang}}]{Li_2018}%
  \BibitemOpen
  \bibfield  {author} {\bibinfo {author} {\bibfnamefont {N.}~\bibnamefont
  {Li}}, \bibinfo {author} {\bibfnamefont {D.}~\bibnamefont {Lu}}, \bibinfo
  {author} {\bibfnamefont {L.}~\bibnamefont {Yang}}, \bibinfo {author}
  {\bibfnamefont {H.}~\bibnamefont {Tao}}, \bibinfo {author} {\bibfnamefont
  {Y.}~\bibnamefont {Xu}}, \bibinfo {author} {\bibfnamefont {C.}~\bibnamefont
  {Wang}}, \bibinfo {author} {\bibfnamefont {L.}~\bibnamefont {Fu}}, \bibinfo
  {author} {\bibfnamefont {H.}~\bibnamefont {Liu}}, \bibinfo {author}
  {\bibfnamefont {Y.}~\bibnamefont {Chummum}}, \ and\ \bibinfo {author}
  {\bibfnamefont {S.}~\bibnamefont {Zhang}},\ }\href {\doibase
  10.1097/aln.0000000000002226} {\bibfield  {journal} {\bibinfo  {journal}
  {Anesthesiology}\ }\textbf {\bibinfo {volume} {129}},\ \bibinfo {pages} {271}
  (\bibinfo {year} {2018})}\BibitemShut {NoStop}%
\bibitem [{\citenamefont {Steiner}\ and\ \citenamefont
  {Ulrich}(1989)}]{Steiner:1989tt}%
  \BibitemOpen
  \bibfield  {author} {\bibinfo {author} {\bibfnamefont {U.~E.}\ \bibnamefont
  {Steiner}}\ and\ \bibinfo {author} {\bibfnamefont {T.}~\bibnamefont
  {Ulrich}},\ }\href {\doibase 10.1021/cr00091a003} {\bibfield  {journal}
  {\bibinfo  {journal} {Chemical Reviews}\ }\textbf {\bibinfo {volume} {89}},\
  \bibinfo {pages} {51} (\bibinfo {year} {1989})}\BibitemShut {NoStop}%
\bibitem [{\citenamefont {Marletto}\ and\ \citenamefont
  {Vedral}(2017)}]{PhysRevLett.119.240402}%
  \BibitemOpen
  \bibfield  {author} {\bibinfo {author} {\bibfnamefont {C.}~\bibnamefont
  {Marletto}}\ and\ \bibinfo {author} {\bibfnamefont {V.}~\bibnamefont
  {Vedral}},\ }\href {\doibase 10.1103/PhysRevLett.119.240402} {\bibfield
  {journal} {\bibinfo  {journal} {Phys. Rev. Lett.}\ }\textbf {\bibinfo
  {volume} {119}},\ \bibinfo {pages} {240402} (\bibinfo {year}
  {2017})}\BibitemShut {NoStop}%
\bibitem [{\citenamefont {Bose}\ \emph {et~al.}(2017)\citenamefont {Bose},
  \citenamefont {Mazumdar}, \citenamefont {Morley}, \citenamefont {Ulbricht},
  \citenamefont {Toro\ifmmode~\check{s}\else \v{s}\fi{}}, \citenamefont
  {Paternostro}, \citenamefont {Geraci}, \citenamefont {Barker}, \citenamefont
  {Kim},\ and\ \citenamefont {Milburn}}]{PhysRevLett.119.240401}%
  \BibitemOpen
  \bibfield  {author} {\bibinfo {author} {\bibfnamefont {S.}~\bibnamefont
  {Bose}}, \bibinfo {author} {\bibfnamefont {A.}~\bibnamefont {Mazumdar}},
  \bibinfo {author} {\bibfnamefont {G.~W.}\ \bibnamefont {Morley}}, \bibinfo
  {author} {\bibfnamefont {H.}~\bibnamefont {Ulbricht}}, \bibinfo {author}
  {\bibfnamefont {M.}~\bibnamefont {Toro\ifmmode~\check{s}\else \v{s}\fi{}}},
  \bibinfo {author} {\bibfnamefont {M.}~\bibnamefont {Paternostro}}, \bibinfo
  {author} {\bibfnamefont {A.~A.}\ \bibnamefont {Geraci}}, \bibinfo {author}
  {\bibfnamefont {P.~F.}\ \bibnamefont {Barker}}, \bibinfo {author}
  {\bibfnamefont {M.~S.}\ \bibnamefont {Kim}}, \ and\ \bibinfo {author}
  {\bibfnamefont {G.}~\bibnamefont {Milburn}},\ }\href {\doibase
  10.1103/PhysRevLett.119.240401} {\bibfield  {journal} {\bibinfo  {journal}
  {Phys. Rev. Lett.}\ }\textbf {\bibinfo {volume} {119}},\ \bibinfo {pages}
  {240401} (\bibinfo {year} {2017})}\BibitemShut {NoStop}%
\bibitem [{\citenamefont {Biham}\ \emph {et~al.}(2004)\citenamefont {Biham},
  \citenamefont {Brassard}, \citenamefont {Kenigsberg},\ and\ \citenamefont
  {Mor}}]{BIHAM200415}%
  \BibitemOpen
  \bibfield  {author} {\bibinfo {author} {\bibfnamefont {E.}~\bibnamefont
  {Biham}}, \bibinfo {author} {\bibfnamefont {G.}~\bibnamefont {Brassard}},
  \bibinfo {author} {\bibfnamefont {D.}~\bibnamefont {Kenigsberg}}, \ and\
  \bibinfo {author} {\bibfnamefont {T.}~\bibnamefont {Mor}},\ }\href {\doibase
  https://doi.org/10.1016/j.tcs.2004.03.041} {\bibfield  {journal} {\bibinfo
  {journal} {Theoretical Computer Science}\ }\textbf {\bibinfo {volume}
  {320}},\ \bibinfo {pages} {15} (\bibinfo {year} {2004})}\BibitemShut
  {NoStop}%
\bibitem [{\citenamefont {G\"arttner}\ \emph {et~al.}(2018)\citenamefont
  {G\"arttner}, \citenamefont {Hauke},\ and\ \citenamefont
  {Rey}}]{PhysRevLett.120.040402}%
  \BibitemOpen
  \bibfield  {author} {\bibinfo {author} {\bibfnamefont {M.}~\bibnamefont
  {G\"arttner}}, \bibinfo {author} {\bibfnamefont {P.}~\bibnamefont {Hauke}}, \
  and\ \bibinfo {author} {\bibfnamefont {A.~M.}\ \bibnamefont {Rey}},\ }\href
  {\doibase 10.1103/PhysRevLett.120.040402} {\bibfield  {journal} {\bibinfo
  {journal} {Phys. Rev. Lett.}\ }\textbf {\bibinfo {volume} {120}},\ \bibinfo
  {pages} {040402} (\bibinfo {year} {2018})}\BibitemShut {NoStop}%
\bibitem [{\citenamefont {Warren}\ \emph {et~al.}(1993)\citenamefont {Warren},
  \citenamefont {Richter}, \citenamefont {Andreotti},\ and\ \citenamefont
  {Farmer}}]{225062:5067241}%
  \BibitemOpen
  \bibfield  {author} {\bibinfo {author} {\bibfnamefont {W.~S.}\ \bibnamefont
  {Warren}}, \bibinfo {author} {\bibfnamefont {W.}~\bibnamefont {Richter}},
  \bibinfo {author} {\bibfnamefont {A.~H.}\ \bibnamefont {Andreotti}}, \ and\
  \bibinfo {author} {\bibfnamefont {B.~T.}\ \bibnamefont {Farmer}},\
  }\href@noop {} {\bibfield  {journal} {\bibinfo  {journal} {Science}\ }\textbf
  {\bibinfo {volume} {262}},\ \bibinfo {pages} {2005} (\bibinfo {year}
  {1993})}\BibitemShut {NoStop}%
\bibitem [{\citenamefont {Jeener}(2000)}]{Jeener_2000}%
  \BibitemOpen
  \bibfield  {author} {\bibinfo {author} {\bibfnamefont {J.}~\bibnamefont
  {Jeener}},\ }\href {\doibase 10.1063/1.481063} {\bibfield  {journal}
  {\bibinfo  {journal} {The Journal of Chemical Physics}\ }\textbf {\bibinfo
  {volume} {112}},\ \bibinfo {pages} {5091} (\bibinfo {year}
  {2000})}\BibitemShut {NoStop}%
\bibitem [{\citenamefont {Deville}\ \emph {et~al.}(1979)\citenamefont
  {Deville}, \citenamefont {Bernier},\ and\ \citenamefont
  {Delrieux}}]{225062:5039968}%
  \BibitemOpen
  \bibfield  {author} {\bibinfo {author} {\bibfnamefont {G.}~\bibnamefont
  {Deville}}, \bibinfo {author} {\bibfnamefont {M.}~\bibnamefont {Bernier}}, \
  and\ \bibinfo {author} {\bibfnamefont {J.~M.}\ \bibnamefont {Delrieux}},\
  }\href@noop {} {\bibfield  {journal} {\bibinfo  {journal} {Physical Review
  B}\ } (\bibinfo {year} {1979})}\BibitemShut {NoStop}%
\bibitem [{\citenamefont {Bowtell}\ \emph {et~al.}(1990)\citenamefont
  {Bowtell}, \citenamefont {Bowley},\ and\ \citenamefont
  {Glover}}]{Bowtell1990}%
  \BibitemOpen
  \bibfield  {author} {\bibinfo {author} {\bibfnamefont {R.}~\bibnamefont
  {Bowtell}}, \bibinfo {author} {\bibfnamefont {R.~M.}\ \bibnamefont {Bowley}},
  \ and\ \bibinfo {author} {\bibfnamefont {P.}~\bibnamefont {Glover}},\
  }\href@noop {} {\bibfield  {journal} {\bibinfo  {journal} {Journal of
  Magnetic Resonance (1969)}\ }\textbf {\bibinfo {volume} {88}},\ \bibinfo
  {pages} {643} (\bibinfo {year} {1990})}\BibitemShut {NoStop}%
\bibitem [{\citenamefont {Bouchard}\ \emph {et~al.}(2002)\citenamefont
  {Bouchard}, \citenamefont {Rizi},\ and\ \citenamefont
  {Warren}}]{225062:5067412}%
  \BibitemOpen
  \bibfield  {author} {\bibinfo {author} {\bibfnamefont {L.~S.}\ \bibnamefont
  {Bouchard}}, \bibinfo {author} {\bibfnamefont {R.}~\bibnamefont {Rizi}}, \
  and\ \bibinfo {author} {\bibfnamefont {W.}~\bibnamefont {Warren}},\
  }\href@noop {} {\bibfield  {journal} {\bibinfo  {journal} {Magn Reson Med}\
  }\textbf {\bibinfo {volume} {48}},\ \bibinfo {pages} {973} (\bibinfo {year}
  {2002})}\BibitemShut {NoStop}%
\bibitem [{\citenamefont {Bowtell}\ \emph {et~al.}(2001)\citenamefont
  {Bowtell}, \citenamefont {Gutteridge},\ and\ \citenamefont
  {Ramanathan}}]{225062:5041128}%
  \BibitemOpen
  \bibfield  {author} {\bibinfo {author} {\bibfnamefont {R.}~\bibnamefont
  {Bowtell}}, \bibinfo {author} {\bibfnamefont {S.}~\bibnamefont {Gutteridge}},
  \ and\ \bibinfo {author} {\bibfnamefont {C.}~\bibnamefont {Ramanathan}},\
  }\href@noop {} {\bibfield  {journal} {\bibinfo  {journal} {Journal of
  Magnetic Resonance}\ }\textbf {\bibinfo {volume} {150}},\ \bibinfo {pages}
  {147} (\bibinfo {year} {2001})}\BibitemShut {NoStop}%
\bibitem [{\citenamefont {Capuani}\ \emph {et~al.}(2004)\citenamefont
  {Capuani}, \citenamefont {Alesiani}, \citenamefont {Branca},\ and\
  \citenamefont {Maraviglia}}]{225062:5067413}%
  \BibitemOpen
  \bibfield  {author} {\bibinfo {author} {\bibfnamefont {S.}~\bibnamefont
  {Capuani}}, \bibinfo {author} {\bibfnamefont {M.}~\bibnamefont {Alesiani}},
  \bibinfo {author} {\bibfnamefont {R.~T.}\ \bibnamefont {Branca}}, \ and\
  \bibinfo {author} {\bibfnamefont {B.}~\bibnamefont {Maraviglia}},\
  }\href@noop {} {\bibfield  {journal} {\bibinfo  {journal} {Solid State
  Nuclear Magnetic Resonance}\ }\textbf {\bibinfo {volume} {25}},\ \bibinfo
  {pages} {153} (\bibinfo {year} {2004})}\BibitemShut {NoStop}%
\bibitem [{\citenamefont {Bouchard}\ \emph {et~al.}(2005)\citenamefont
  {Bouchard}, \citenamefont {Wehrli}, \citenamefont {Chin},\ and\ \citenamefont
  {Warren}}]{225062:5067414}%
  \BibitemOpen
  \bibfield  {author} {\bibinfo {author} {\bibfnamefont {L.~S.}\ \bibnamefont
  {Bouchard}}, \bibinfo {author} {\bibfnamefont {F.~W.}\ \bibnamefont
  {Wehrli}}, \bibinfo {author} {\bibfnamefont {C.~L.}\ \bibnamefont {Chin}}, \
  and\ \bibinfo {author} {\bibfnamefont {S.~W.}\ \bibnamefont {Warren}},\
  }\href@noop {} {\bibfield  {journal} {\bibinfo  {journal} {Journal of
  Magnetic Resonance}\ }\textbf {\bibinfo {volume} {176}},\ \bibinfo {pages}
  {27} (\bibinfo {year} {2005})}\BibitemShut {NoStop}%
\bibitem [{\citenamefont {Heidrich-Meisner}\ \emph {et~al.}(2009)\citenamefont
  {Heidrich-Meisner}, \citenamefont {Manmana}, \citenamefont {Rigol},
  \citenamefont {Muramatsu}, \citenamefont {Feiguin},\ and\ \citenamefont
  {Dagotto}}]{PhysRevA.80.041603}%
  \BibitemOpen
  \bibfield  {author} {\bibinfo {author} {\bibfnamefont {F.}~\bibnamefont
  {Heidrich-Meisner}}, \bibinfo {author} {\bibfnamefont {S.~R.}\ \bibnamefont
  {Manmana}}, \bibinfo {author} {\bibfnamefont {M.}~\bibnamefont {Rigol}},
  \bibinfo {author} {\bibfnamefont {A.}~\bibnamefont {Muramatsu}}, \bibinfo
  {author} {\bibfnamefont {A.~E.}\ \bibnamefont {Feiguin}}, \ and\ \bibinfo
  {author} {\bibfnamefont {E.}~\bibnamefont {Dagotto}},\ }\href {\doibase
  10.1103/PhysRevA.80.041603} {\bibfield  {journal} {\bibinfo  {journal} {Phys.
  Rev. A}\ }\textbf {\bibinfo {volume} {80}},\ \bibinfo {pages} {041603}
  (\bibinfo {year} {2009})}\BibitemShut {NoStop}%
\bibitem [{\citenamefont {Bravyi}\ and\ \citenamefont
  {Kitaev}(2005)}]{PhysRevA.71.022316}%
  \BibitemOpen
  \bibfield  {author} {\bibinfo {author} {\bibfnamefont {S.}~\bibnamefont
  {Bravyi}}\ and\ \bibinfo {author} {\bibfnamefont {A.}~\bibnamefont
  {Kitaev}},\ }\href {\doibase 10.1103/PhysRevA.71.022316} {\bibfield
  {journal} {\bibinfo  {journal} {Phys. Rev. A}\ }\textbf {\bibinfo {volume}
  {71}},\ \bibinfo {pages} {022316} (\bibinfo {year} {2005})}\BibitemShut
  {NoStop}%
\bibitem [{\citenamefont {Braunstein}\ \emph {et~al.}(1999)\citenamefont
  {Braunstein}, \citenamefont {Caves}, \citenamefont {Jozsa}, \citenamefont
  {Linden}, \citenamefont {Popescu},\ and\ \citenamefont
  {Schack}}]{PhysRevLett.83.1054}%
  \BibitemOpen
  \bibfield  {author} {\bibinfo {author} {\bibfnamefont {S.~L.}\ \bibnamefont
  {Braunstein}}, \bibinfo {author} {\bibfnamefont {C.~M.}\ \bibnamefont
  {Caves}}, \bibinfo {author} {\bibfnamefont {R.}~\bibnamefont {Jozsa}},
  \bibinfo {author} {\bibfnamefont {N.}~\bibnamefont {Linden}}, \bibinfo
  {author} {\bibfnamefont {S.}~\bibnamefont {Popescu}}, \ and\ \bibinfo
  {author} {\bibfnamefont {R.}~\bibnamefont {Schack}},\ }\href {\doibase
  10.1103/PhysRevLett.83.1054} {\bibfield  {journal} {\bibinfo  {journal}
  {Phys. Rev. Lett.}\ }\textbf {\bibinfo {volume} {83}},\ \bibinfo {pages}
  {1054} (\bibinfo {year} {1999})}\BibitemShut {NoStop}%
\bibitem [{\citenamefont {Wie{\'s}niak}\ \emph {et~al.}(2005)\citenamefont
  {Wie{\'s}niak}, \citenamefont {Vedral},\ and\ \citenamefont
  {Brukner}}]{10.1088/1367-2630/7/1/258}%
  \BibitemOpen
  \bibfield  {author} {\bibinfo {author} {\bibfnamefont {M.}~\bibnamefont
  {Wie{\'s}niak}}, \bibinfo {author} {\bibfnamefont {V.}~\bibnamefont
  {Vedral}}, \ and\ \bibinfo {author} {\bibfnamefont {{\v C}.}~\bibnamefont
  {Brukner}},\ }\href {\doibase 10.1088/1367-2630/7/1/258} {\bibfield
  {journal} {\bibinfo  {journal} {New Journal of Physics}\ }\textbf {\bibinfo
  {volume} {7}},\ \bibinfo {pages} {258} (\bibinfo {year} {2005})}\BibitemShut
  {NoStop}%
\bibitem [{\citenamefont {Fan}\ \emph {et~al.}(2019)\citenamefont {Fan},
  \citenamefont {Sun}, \citenamefont {Ding}, \citenamefont {Ming},
  \citenamefont {Yang}, \citenamefont {Wang},\ and\ \citenamefont
  {Ye}}]{Fan_2019}%
  \BibitemOpen
  \bibfield  {author} {\bibinfo {author} {\bibfnamefont {X.-G.}\ \bibnamefont
  {Fan}}, \bibinfo {author} {\bibfnamefont {W.-Y.}\ \bibnamefont {Sun}},
  \bibinfo {author} {\bibfnamefont {Z.-Y.}\ \bibnamefont {Ding}}, \bibinfo
  {author} {\bibfnamefont {F.}~\bibnamefont {Ming}}, \bibinfo {author}
  {\bibfnamefont {H.}~\bibnamefont {Yang}}, \bibinfo {author} {\bibfnamefont
  {D.}~\bibnamefont {Wang}}, \ and\ \bibinfo {author} {\bibfnamefont
  {L.}~\bibnamefont {Ye}},\ }\href {\doibase 10.1088/1367-2630/ab41b1}
  {\bibfield  {journal} {\bibinfo  {journal} {New Journal of Physics}\ }\textbf
  {\bibinfo {volume} {21}},\ \bibinfo {pages} {093053} (\bibinfo {year}
  {2019})}\BibitemShut {NoStop}%
\bibitem [{\citenamefont {Park}\ and\ \citenamefont
  {Blanke}(2019)}]{PARK2019502}%
  \BibitemOpen
  \bibfield  {author} {\bibinfo {author} {\bibfnamefont {H.-D.}\ \bibnamefont
  {Park}}\ and\ \bibinfo {author} {\bibfnamefont {O.}~\bibnamefont {Blanke}},\
  }\href {\doibase https://doi.org/10.1016/j.neuroimage.2019.04.081} {\bibfield
   {journal} {\bibinfo  {journal} {NeuroImage}\ }\textbf {\bibinfo {volume}
  {197}},\ \bibinfo {pages} {502} (\bibinfo {year} {2019})}\BibitemShut
  {NoStop}%
\bibitem [{\citenamefont {Zhong}\ \emph {et~al.}(2001)\citenamefont {Zhong},
  \citenamefont {Shaokuan},\ and\ \citenamefont {Jianhui}}]{225062:5277829}%
  \BibitemOpen
  \bibfield  {author} {\bibinfo {author} {\bibfnamefont {C.}~\bibnamefont
  {Zhong}}, \bibinfo {author} {\bibfnamefont {Z.}~\bibnamefont {Shaokuan}}, \
  and\ \bibinfo {author} {\bibfnamefont {Z.}~\bibnamefont {Jianhui}},\
  }\href@noop {} {\bibfield  {journal} {\bibinfo  {journal} {Chemical Physics
  Letters}\ }\textbf {\bibinfo {volume} {347}},\ \bibinfo {pages} {143}
  (\bibinfo {year} {2001})}\BibitemShut {NoStop}%
\bibitem [{\citenamefont {Ernst}\ and\ \citenamefont
  {Anderson}(1966)}]{Ernst_1966}%
  \BibitemOpen
  \bibfield  {author} {\bibinfo {author} {\bibfnamefont {R.~R.}\ \bibnamefont
  {Ernst}}\ and\ \bibinfo {author} {\bibfnamefont {W.~A.}\ \bibnamefont
  {Anderson}},\ }\href {\doibase 10.1063/1.1719961} {\bibfield  {journal}
  {\bibinfo  {journal} {Review of Scientific Instruments}\ }\textbf {\bibinfo
  {volume} {37}},\ \bibinfo {pages} {93} (\bibinfo {year} {1966})}\BibitemShut
  {NoStop}%
\bibitem [{\citenamefont {Uzi}\ and\ \citenamefont
  {Gil}(2008)}]{225062:5277825}%
  \BibitemOpen
  \bibfield  {author} {\bibinfo {author} {\bibfnamefont {E.}~\bibnamefont
  {Uzi}}\ and\ \bibinfo {author} {\bibfnamefont {N.}~\bibnamefont {Gil}},\
  }\href@noop {} {\bibfield  {journal} {\bibinfo  {journal} {Journal of
  Magnetic Resonance}\ }\textbf {\bibinfo {volume} {190}},\ \bibinfo {pages}
  {149} (\bibinfo {year} {2008})}\BibitemShut {NoStop}%
\bibitem [{\citenamefont {Nunes}\ \emph {et~al.}(2005)\citenamefont {Nunes},
  \citenamefont {Jezzard},\ and\ \citenamefont {Clare}}]{225062:5067666}%
  \BibitemOpen
  \bibfield  {author} {\bibinfo {author} {\bibfnamefont {R.}~\bibnamefont
  {Nunes}}, \bibinfo {author} {\bibfnamefont {P.}~\bibnamefont {Jezzard}}, \
  and\ \bibinfo {author} {\bibfnamefont {S.}~\bibnamefont {Clare}},\
  }\href@noop {} {\bibfield  {journal} {\bibinfo  {journal} {J Magn Reson}\
  }\textbf {\bibinfo {volume} {177}},\ \bibinfo {pages} {102} (\bibinfo {year}
  {2005})}\BibitemShut {NoStop}%
\bibitem [{\citenamefont {Marwan}\ and\ \citenamefont
  {Kurths}(2002)}]{MARWAN2002299}%
  \BibitemOpen
  \bibfield  {author} {\bibinfo {author} {\bibfnamefont {N.}~\bibnamefont
  {Marwan}}\ and\ \bibinfo {author} {\bibfnamefont {J.}~\bibnamefont
  {Kurths}},\ }\href {\doibase https://doi.org/10.1016/S0375-9601(02)01170-2}
  {\bibfield  {journal} {\bibinfo  {journal} {Physics Letters A}\ }\textbf
  {\bibinfo {volume} {302}},\ \bibinfo {pages} {299} (\bibinfo {year}
  {2002})}\BibitemShut {NoStop}%
\bibitem [{web()}]{webpage}%
  \BibitemOpen
  \href {http://tocsy.pik-potsdam.de/CRPtoolbox/} {}\BibitemShut {NoStop}%
\bibitem [{\citenamefont {Marwan}\ \emph {et~al.}(2007)\citenamefont {Marwan},
  \citenamefont {{Carmen Romano}}, \citenamefont {Thiel},\ and\ \citenamefont
  {Kurths}}]{MARWAN2007237}%
  \BibitemOpen
  \bibfield  {author} {\bibinfo {author} {\bibfnamefont {N.}~\bibnamefont
  {Marwan}}, \bibinfo {author} {\bibfnamefont {M.}~\bibnamefont {{Carmen
  Romano}}}, \bibinfo {author} {\bibfnamefont {M.}~\bibnamefont {Thiel}}, \
  and\ \bibinfo {author} {\bibfnamefont {J.}~\bibnamefont {Kurths}},\ }\href
  {\doibase https://doi.org/10.1016/j.physrep.2006.11.001} {\bibfield
  {journal} {\bibinfo  {journal} {Physics Reports}\ }\textbf {\bibinfo {volume}
  {438}},\ \bibinfo {pages} {237} (\bibinfo {year} {2007})}\BibitemShut
  {NoStop}%
\bibitem [{\citenamefont {MATLAB}(2014)}]{MATLAB:2014}%
  \BibitemOpen
  \bibfield  {author} {\bibinfo {author} {\bibnamefont {MATLAB}},\ }\href@noop
  {} {\emph {\bibinfo {title} {version 2014a}}}\ (\bibinfo  {publisher} {The
  MathWorks Inc.},\ \bibinfo {address} {Natick, Massachusetts},\ \bibinfo
  {year} {2014})\BibitemShut {NoStop}%
\bibitem [{\citenamefont {Gomes}\ \emph {et~al.}(2013)\citenamefont {Gomes},
  \citenamefont {Jorge},\ and\ \citenamefont {Azevedo}}]{225062:5042988}%
  \BibitemOpen
  \bibfield  {author} {\bibinfo {author} {\bibfnamefont {E.~F.}\ \bibnamefont
  {Gomes}}, \bibinfo {author} {\bibfnamefont {A.~M.}\ \bibnamefont {Jorge}}, \
  and\ \bibinfo {author} {\bibfnamefont {P.~J.}\ \bibnamefont {Azevedo}},\ }in\
  \href@noop {} {\emph {\bibinfo {booktitle} {Proceedings of the International
  C* Conference on Computer Science and Software Engineering}}}\ (\bibinfo
  {publisher} {ACM},\ \bibinfo {year} {2013})\ pp.\ \bibinfo {pages}
  {23--30}\BibitemShut {NoStop}%
\bibitem [{\citenamefont {L{\'o}pez~P{\'e}rez}\ \emph
  {et~al.}(2020)\citenamefont {L{\'o}pez~P{\'e}rez}, \citenamefont {Bokde},\
  and\ \citenamefont {Kerskens}}]{Perez:2020aa}%
  \BibitemOpen
  \bibfield  {author} {\bibinfo {author} {\bibfnamefont {D.}~\bibnamefont
  {L{\'o}pez~P{\'e}rez}}, \bibinfo {author} {\bibfnamefont {A.~L.~W.}\
  \bibnamefont {Bokde}}, \ and\ \bibinfo {author} {\bibfnamefont
  {C.}~\bibnamefont {Kerskens}},\ }\href {\doibase 10.1101/2020.05.27.117226}
  {\bibfield  {journal} {\bibinfo  {journal} {recommended for publication in
  EPJ}\ ,\ \bibinfo {pages} {2020.05.27.117226}} (\bibinfo {year}
  {2020})}\BibitemShut {NoStop}%
\bibitem [{\citenamefont {Inc.}()}]{Mathematica}%
  \BibitemOpen
  \bibfield  {author} {\bibinfo {author} {\bibfnamefont {W.~R.}\ \bibnamefont
  {Inc.}},\ }\href {https://www.wolfram.com/mathematica} {\enquote {\bibinfo
  {title} {Mathematica, {V}ersion 12.3.1},}\ }\bibinfo {note} {Champaign, IL,
  2021}\BibitemShut {NoStop}%
\bibitem [{\citenamefont {Dagli}\ \emph {et~al.}(1999)\citenamefont {Dagli},
  \citenamefont {Ingeholm},\ and\ \citenamefont {Haxby}}]{DAGLI1999407}%
  \BibitemOpen
  \bibfield  {author} {\bibinfo {author} {\bibfnamefont {M.~S.}\ \bibnamefont
  {Dagli}}, \bibinfo {author} {\bibfnamefont {J.~E.}\ \bibnamefont {Ingeholm}},
  \ and\ \bibinfo {author} {\bibfnamefont {J.~V.}\ \bibnamefont {Haxby}},\
  }\href {\doibase https://doi.org/10.1006/nimg.1998.0424} {\bibfield
  {journal} {\bibinfo  {journal} {NeuroImage}\ }\textbf {\bibinfo {volume}
  {9}},\ \bibinfo {pages} {407} (\bibinfo {year} {1999})}\BibitemShut {NoStop}%
\bibitem [{\citenamefont {L{\'o}pez~P{\'e}rez}(2015)}]{davidprivate}%
  \BibitemOpen
  \bibfield  {author} {\bibinfo {author} {\bibfnamefont {D.}~\bibnamefont
  {L{\'o}pez~P{\'e}rez}},\ }\href@noop {} {\enquote {\bibinfo {title} {Private
  communication},}\ } (\bibinfo {year} {2015})\BibitemShut {NoStop}%
\bibitem [{\citenamefont {Baum}\ \emph {et~al.}(1985)\citenamefont {Baum},
  \citenamefont {Munowitz}, \citenamefont {Garroway},\ and\ \citenamefont
  {Pines}}]{225062:5187628}%
  \BibitemOpen
  \bibfield  {author} {\bibinfo {author} {\bibfnamefont {J.}~\bibnamefont
  {Baum}}, \bibinfo {author} {\bibfnamefont {M.}~\bibnamefont {Munowitz}},
  \bibinfo {author} {\bibfnamefont {A.~N.}\ \bibnamefont {Garroway}}, \ and\
  \bibinfo {author} {\bibfnamefont {A.}~\bibnamefont {Pines}},\ }\href@noop {}
  {\bibfield  {journal} {\bibinfo  {journal} {Multiple-quantum dynamics in
  solid state NMR,J. Chem. Phys. 83, 2015 (1985)}\ } (\bibinfo {year}
  {1985})}\BibitemShut {NoStop}%
\bibitem [{\citenamefont {Nakamura}\ \emph {et~al.}(2009)\citenamefont
  {Nakamura}, \citenamefont {Miyati}, \citenamefont {Kasai}, \citenamefont
  {Ohno}, \citenamefont {Yamada}, \citenamefont {Mase}, \citenamefont {Hara},
  \citenamefont {Shibamoto}, \citenamefont {Suzuki},\ and\ \citenamefont
  {Ichikawa}}]{Nakamura:2009uq}%
  \BibitemOpen
  \bibfield  {author} {\bibinfo {author} {\bibfnamefont {T.}~\bibnamefont
  {Nakamura}}, \bibinfo {author} {\bibfnamefont {T.}~\bibnamefont {Miyati}},
  \bibinfo {author} {\bibfnamefont {H.}~\bibnamefont {Kasai}}, \bibinfo
  {author} {\bibfnamefont {N.}~\bibnamefont {Ohno}}, \bibinfo {author}
  {\bibfnamefont {M.}~\bibnamefont {Yamada}}, \bibinfo {author} {\bibfnamefont
  {M.}~\bibnamefont {Mase}}, \bibinfo {author} {\bibfnamefont {M.}~\bibnamefont
  {Hara}}, \bibinfo {author} {\bibfnamefont {Y.}~\bibnamefont {Shibamoto}},
  \bibinfo {author} {\bibfnamefont {Y.}~\bibnamefont {Suzuki}}, \ and\ \bibinfo
  {author} {\bibfnamefont {K.}~\bibnamefont {Ichikawa}},\ }\href {\doibase
  10.1007/s12194-009-0056-3} {\bibfield  {journal} {\bibinfo  {journal}
  {Radiological Physics and Technology}\ }\textbf {\bibinfo {volume} {2}},\
  \bibinfo {pages} {133} (\bibinfo {year} {2009})}\BibitemShut {NoStop}%
\bibitem [{\citenamefont {Warren}\ \emph {et~al.}(1998)\citenamefont {Warren},
  \citenamefont {Ahn}, \citenamefont {Mescher}, \citenamefont {Garwood},
  \citenamefont {Ugurbil}, \citenamefont {Richter}, \citenamefont {Rizi},
  \citenamefont {Hopkins},\ and\ \citenamefont {Leigh}}]{225062:5289250}%
  \BibitemOpen
  \bibfield  {author} {\bibinfo {author} {\bibfnamefont {W.~S.}\ \bibnamefont
  {Warren}}, \bibinfo {author} {\bibfnamefont {S.}~\bibnamefont {Ahn}},
  \bibinfo {author} {\bibfnamefont {M.}~\bibnamefont {Mescher}}, \bibinfo
  {author} {\bibfnamefont {M.}~\bibnamefont {Garwood}}, \bibinfo {author}
  {\bibfnamefont {K.}~\bibnamefont {Ugurbil}}, \bibinfo {author} {\bibfnamefont
  {W.}~\bibnamefont {Richter}}, \bibinfo {author} {\bibfnamefont {R.~R.}\
  \bibnamefont {Rizi}}, \bibinfo {author} {\bibfnamefont {J.}~\bibnamefont
  {Hopkins}}, \ and\ \bibinfo {author} {\bibfnamefont {J.~S.}\ \bibnamefont
  {Leigh}},\ }\href@noop {} {\bibfield  {journal} {\bibinfo  {journal}
  {Science}\ }\textbf {\bibinfo {volume} {281}},\ \bibinfo {pages} {247}
  (\bibinfo {year} {1998})}\BibitemShut {NoStop}%
\bibitem [{\citenamefont {Rizi}\ \emph {et~al.}(2000)\citenamefont {Rizi},
  \citenamefont {Ahn}, \citenamefont {Alsop}, \citenamefont {Garrett-Roe},
  \citenamefont {Mescher}, \citenamefont {Richter}, \citenamefont {Schnall},
  \citenamefont {Leigh},\ and\ \citenamefont {Warren}}]{Rizi_2000}%
  \BibitemOpen
  \bibfield  {author} {\bibinfo {author} {\bibfnamefont {R.~R.}\ \bibnamefont
  {Rizi}}, \bibinfo {author} {\bibfnamefont {S.}~\bibnamefont {Ahn}}, \bibinfo
  {author} {\bibfnamefont {D.~C.}\ \bibnamefont {Alsop}}, \bibinfo {author}
  {\bibfnamefont {S.}~\bibnamefont {Garrett-Roe}}, \bibinfo {author}
  {\bibfnamefont {M.}~\bibnamefont {Mescher}}, \bibinfo {author} {\bibfnamefont
  {W.}~\bibnamefont {Richter}}, \bibinfo {author} {\bibfnamefont {M.~D.}\
  \bibnamefont {Schnall}}, \bibinfo {author} {\bibfnamefont {J.~S.}\
  \bibnamefont {Leigh}}, \ and\ \bibinfo {author} {\bibfnamefont {W.~S.}\
  \bibnamefont {Warren}},\ }\href {\doibase
  10.1002/(sici)1522-2594(200005)43:5<627::aid-mrm2>3.0.co;2-j} {\bibfield
  {journal} {\bibinfo  {journal} {Magnetic Resonance in Medicine}\ }\textbf
  {\bibinfo {volume} {43}},\ \bibinfo {pages} {627} (\bibinfo {year}
  {2000})}\BibitemShut {NoStop}%
\bibitem [{\citenamefont {Kern}\ \emph {et~al.}(2013)\citenamefont {Kern},
  \citenamefont {Aertsen}, \citenamefont {Schulze-Bonhage},\ and\ \citenamefont
  {Ball}}]{KERN2013178}%
  \BibitemOpen
  \bibfield  {author} {\bibinfo {author} {\bibfnamefont {M.}~\bibnamefont
  {Kern}}, \bibinfo {author} {\bibfnamefont {A.}~\bibnamefont {Aertsen}},
  \bibinfo {author} {\bibfnamefont {A.}~\bibnamefont {Schulze-Bonhage}}, \ and\
  \bibinfo {author} {\bibfnamefont {T.}~\bibnamefont {Ball}},\ }\href {\doibase
  https://doi.org/10.1016/j.neuroimage.2013.05.042} {\bibfield  {journal}
  {\bibinfo  {journal} {NeuroImage}\ }\textbf {\bibinfo {volume} {81}},\
  \bibinfo {pages} {178} (\bibinfo {year} {2013})}\BibitemShut {NoStop}%
\bibitem [{\citenamefont {Simor}\ \emph {et~al.}(2021)\citenamefont {Simor},
  \citenamefont {Bogd{\'a}ny}, \citenamefont {B{\'o}dizs},\ and\ \citenamefont
  {Perakakis}}]{10.1093/sleep/zsab100}%
  \BibitemOpen
  \bibfield  {author} {\bibinfo {author} {\bibfnamefont {P.}~\bibnamefont
  {Simor}}, \bibinfo {author} {\bibfnamefont {T.}~\bibnamefont {Bogd{\'a}ny}},
  \bibinfo {author} {\bibfnamefont {R.}~\bibnamefont {B{\'o}dizs}}, \ and\
  \bibinfo {author} {\bibfnamefont {P.}~\bibnamefont {Perakakis}},\ }\href
  {\doibase 10.1093/sleep/zsab100} {\bibfield  {journal} {\bibinfo  {journal}
  {Sleep}\ }\textbf {\bibinfo {volume} {44}} (\bibinfo {year} {2021}),\
  10.1093/sleep/zsab100},\ \bibinfo {note} {zsab100},\ \Eprint
  {http://arxiv.org/abs/https://academic.oup.com/sleep/article-pdf/44/9/zsab100/41327815/zsab100.pdf}
  {https://academic.oup.com/sleep/article-pdf/44/9/zsab100/41327815/zsab100.pdf}
  \BibitemShut {NoStop}%
\bibitem [{Note1()}]{Note1}%
  \BibitemOpen
  \bibinfo {note} {Using the conventional MQC sequence design}\BibitemShut
  {NoStop}%
\end{thebibliography}%

\end{document}